\makeatletter
\newcommand{\thickhline}{%
	\noalign {\ifnum 0=`}\fi \hrule height 1pt
	\futurelet \reserved@a \@xhline
}

\documentclass[journal]{IEEEtran}[10pts]
\usepackage{algorithm}
\usepackage{algpseudocode}
\usepackage{tabularx}
\usepackage{graphicx}
\usepackage{caption}
\usepackage{subcaption}
\usepackage{amsmath}
\usepackage{epstopdf}
\usepackage{tabularx}
\usepackage{array}
\usepackage{authblk}
\usepackage{tabularx}
\usepackage{amsmath,graphicx}
\usepackage{mathtools}
\usepackage{amsthm}
\usepackage{epstopdf}
\usepackage{array}
\usepackage{authblk}

\usepackage{relsize}

\title{Multivariate Variational Mode Decomposition}
\author{Naveed ur Rehman and Hania Aftab 
	\thanks{N. Rehman, H. Aftab  are with the Department
		of Electrical Engineering, COMSATS University Islamabad, Islamabad,
		Pakistan e-mail: naveed.rehman@comsats.edu.pk.\\
	\textcopyright2019 IEEE. Personal use of this material is permitted. Permission from
	IEEE must be obtained for all other uses, in any current or future media,
	including reprinting/republishing this material for advertising or promotional
	purposes, creating new collective works, for resale or redistribution to servers
	or lists, or reuse of any copyrighted component of this work in other works.}}
\begin{document}
	%
	\maketitle

\begin{abstract}
In this paper, a generic extension of variational mode decomposition (VMD) algorithm for multivariate or multichannel data sets is presented. We first define a model for multivariate modulated oscillations that is based on the presence of a joint or common frequency component among all channels of input data. Using that model for multivariate oscillations, we construct a variational optimization problem that aims to extract an ensemble of band-limited modes containing inherent multivariate modulated oscillations present in multivariate input signal. The cost function to be minimized is the sum of bandwidths of all signal modes across all input data channels, which is a generic extension of the cost function used in standard VMD to multivariate data. 
Minimization of the resulting variational model is achieved through the alternate direction method of multipliers (ADMM) approach. That yields an optimal set of multivariate modes in terms of narrow bandwidth and corresponding center frequencies that are assumed to be commonly present among all channels of a multivariate modulated oscillation. We demonstrate the effectiveness of the proposed method through results obtained from extensive simulations involving test (synthetic) and real world multivariate data sets. Specifically, we focus on the ability of the proposed method to yield joint oscillatory modes in multivariate data which is a prerequisite in many real world applications involving nonstationary multivariate data. We also highlight the utility of the proposed method in two real world applications which include the separation of alpha rhythms in multivariate electroencephalogram (EEG) data and the decomposition of bivariate cardiotocographic signals that consist of fetal heart rate and maternal uterine contraction (FHR-UC) as its two channels.

\end{abstract}


\section{Introduction}

\IEEEPARstart{N}{onstationary} signal processing methods have attracted a lot of interest over the last few decades owing to their relevance and applicability to a large class of real world signals. Here we are often interested in tracking time variations of packets of close frequencies in input data. One way of achieving that is through linear time-frequency transforms that correlate or project input data with a family of signals, termed as time frequency atoms, that are well concentrated in both time and frequency. Short Time Fourier Transform (STFT) \cite{bib:stft} and wavelet transform \cite{bib:wavelet} are two popular examples of linear transform methods. In addition to providing a graphical display in the form of time-frequency (T-F) spectrum (spectrogram or scalogram) for the purpose of exploratory data analysis, these methods enable signal separation at multiple resolution thus enabling signal processing at each scale separately and subsequent reconstruction.          

A downside of linear transform methods is that the time frequency atoms are fixed and predefined. They exhibit limited joint time and frequency  resolution which blurs the resulting T-F representation due to the famous Heisenberg uncertainty principle. That may be problematic in many real world applications that involve intermittent signal patterns i.e. nonstationarity. To that end, there has been a surge of interest in data-driven signal decomposition and T-F methods that make little or no a priori assumptions on input data. 

The trend started in the late nineties when Huang \textit{et al.} proposed a recursive algorithm, termed empirical mode decomposition (EMD) \cite{bib:emd}. EMD decomposes input data into its inherent oscillatory modes, called as intrinsic mode functions (IMFs), through a recursive sifting process that makes use of signal extrema. Specifically, signal extrema are interpolated to yield upper and lower envelopes which are averaged to obtain a local mean of the signal. Local mean, which are low-frequency estimates of the data, are then recursively removed from input data to yield a high-frequency (or fast oscillating) mode in the signal. The process is repeated until all principal oscillatory modes present in the data are recovered. Owing to its fully data driven nature, EMD avoids several limitations of linear T-F methods, such as limited T-F resolution (Heisenberg uncertainty theorem) and physically meaningless modes obtained due to the use of fixed T-F atoms. Consequently, EMD has made a significant impact in scientific community and is routinely employed in wide-ranging engineering applications such as biomedical signal classification \cite{bib:bio, bib:vag}, climate analysis \cite{bib:clim} and off-line machine monitoring \cite{bib:mach}. 

In essence, the goal of EMD is to extract multicomponent signals (IMFs) $c_k(t)$ and trend $r(t)$, from input data $x(t)$, as follows

\begin{equation}
x(t)=\sum_{k=1}^{K}c_k(t)+r(t),
\label{eq:emd}
\end{equation}  

\noindent where $c_k(t)$ are defined such that they are amplitude- and frequency-modulated (AM-FM) signals

\begin{equation}
c_k(t)=a_k(t)\cos(\phi_k(t)).
\label{eq:am-fm}
\end{equation}  

Here, $\phi_k(t)$ denotes instantaneous phase which satisfies $\phi^{'}(t)\geq 0\mbox{, }\forall t$ and the instantaneous amplitude $a_k(t)$ is nonnegative i.e., $a_k(t)\geq 0 \mbox{, }\forall t$. The instantaneous frequency (IF), denoted by $\omega_k(t)$, is then defined by $\omega_k(t)=\frac{d\phi(t)}{dt}$ and is assumed to vary much slower than the phase $\phi_k(t)$.  
   
Despite its advantages, EMD remains a recursive process whose output depends heavily on the choice of interpolation scheme and stopping criteria to stop the sifting process. Due to the lack of mathematical theory, there is a little or no margin to obtain necessary theoretical guarantees for EMD decomposition.

To overcome that difficulty, earlier attempts included replacing envelope and local mean estimation in EMD with robust constrained optimization based methods \cite{bib:emd_form1, bib:emd_form2}. Another class of data driven methods that aim to extract EMD-like decomposition include synchrosqueezed transform (SST) \cite{bib:sst} and empirical wavelet transform (EWT) \cite{bib:ewt}. SST belongs to a class of methods that obtains AM-FM signal components as given in \eqref{eq:am-fm}, albeit through a sound and convenient mathematical framework in contrast to EMD. It can operate both in the STFT and wavelet domain and works by first sharpening the signals' STFT spectrogram (or wavelet scalogram) through a frequency reassignment operator, similar to the one proposed in \cite{bib:reassign}. Then, assuming that the total number of signal components are known a priori, ridge extraction techniques are employed to estimate ridges of IFs i.e., $\omega_k(t)$. Finally, signal components are recovered by integrating the reassigned STFT (or reassigned wavelet transform) in the vicinity of the corresponding ridges of $\omega_k(t)$. EWT, on the other hand, adopts an empirical approach that relies on robust peak detection mechanism and spectrum segmentation techniques based on the detected peaks to develop a wavelet filter bank for effective data decomposition. 

Recently, a variational approach to data decomposition, termed variational mode decomposition (VMD), has been proposed that obtains principal modes of a signal, similar in form to \eqref{eq:am-fm}, adaptively and concurrently by solving a convex optimization problem \cite{bib:vmd}. In this approach, an optimization problem is formulated in order minimize the collective bandwidth of modes, subject to the constraint that the modes fully reconstruct the input signal. The bandwidth is measured as $\mathcal{H}^1$ norm of corresponding analytic signal shifted to the baseband via complex harmonic mixing. The resulting variational model is minimized through alternate direction method of multipliers (ADMM) approach to find an ensemble of modes that are band-limited around a center frequency $\omega_k$; $\omega_k$ is determined online during the optimization process.   

While efforts to develop fully data driven yet mathematically sound algorithms for signal decomposition and T-F analysis of nonstationary data continue to grow, there has been a lot of interest in extending existing data driven approaches to process nonstationary \textit{multidimensional and multivariate (multichannel)} data sets. Owing to advances in sensor and computing technologies, such data types are routinely acquired and need to be processed in modern engineering and scientific applications e.g., classification based on multichannel electroencephalogram (EEG) \cite{bib:memd_seizure} and ECG signals \cite{bib:memd_ecg} and denoising \cite{bib:memd_den}. The main challenges involved in multivariate T-F algorithms for nonstationary data include: i) alignment of frequency information across multiple channels in each mode, termed as \textit{mode-alignment}; ii) incorporating any correlation between multiple data channels. Typically, developing extensions that operate directly in multidimensional spaces where input signal resides are able to fulfill the above requirements.

Multivariate signal processing using standard linear transform methods is straightforward. Fixed basis functions (filters) associated with these transforms can be applied to each channel of a multivariate signal separately. For fully data driven methods like EMD, however, specialized extensions are employed for multivariate data that operate directly in multidimensional space where signal resides. A generic multivariate extension of EMD, termed as multivariate EMD (MEMD) \cite{bib:memd} belongs to this class of methods that can operate on multichannel data with any number of input channels. While univariate EMD works on the principle of separating faster oscillations from slower oscillations, MEMD aims to separate faster \textit{multivariate oscillations} from slower ones. To accomplish that, local mean of a multivariate signal is directly estimated in multidimensional space using a uniform projection based approach. MEMD and other extensions of EMD, especially for  bivariate \cite{bib:bemd} and trivariate \cite{bib:temd} signals, have been utilized in wide ranging applications involving multivariate data e.g., in biomedical engineering applications involving EEG  and ECG, acoustics, fault detection and machine monitoring based applications to name a few. However, these extensions inherit all the limitations of standard EMD such as sensitivity to sampling rate, lack of robustness to noise and issues related to the empirical and algorithmic nature of the EMD algorithm.              

Multivariate extensions for wavelet based methods such as SST \cite{bib:msst} and EWT \cite{bib:mewt} have also emerged recently. The multivariate extension of SST operates by first applying the standard SST algorithm to each channel separately. That is followed by an adaptive partitioning of the T-F domain in order to separate monocomponent multivariate oscillations in the input data. Finally, multivariate instantaneous frequency and amplitude are estimated and a multivariate synchrosqueezed transform operator is computed based on those estimates. That results in a sharp T-F spectrum of multivariate data which is useful for exploratory data analysis. However, individual modes cannot be reconstructed using this approach which limits its applications. The multivariate extension of EWT \cite{bib:mewt} employs a mode estimation procedure to obtain an optimum signal in the multivariate data set and then segments its corresponding spectra to recover relevant modes across all the channels of input data. The method suffers from inherent flaws of EWT in that it requires explicit construction of adaptive wavelet filters based on spectrum segmentation.  

Recently, an extension of VMD to process complex-valued signals has been proposed \cite{bib:cvmd}. The paper investigates filterbank structure of complex VMD for wGn and also proposes bi-directional T-F spectrum for complex-valued data based on complex VMD. The method, by design, is however not suitable for extracting matched frequency components across its channels  \cite{bib:looney, bib:memd}. In addition, it is applicable only for bivariate data containing up to two channels only.         

In this paper, we present a novel extension of the VMD algorithm, termed multivariate VMD (MVMD), to process multivariate data containing any number of channels. This is the first fully multivariate extension of VMD and is important due to the following limitations of existing multivariate data driven approaches: i) multichannel extension of EMD, namely MEMD, suffers from all the problems of standard EMD, including absence of mathematical framework and lack of robustness to low sampling rates and noise. mode-mixing among different channels of MEMD \cite{bib:fb_memd} is also problematic; ii) multivariate extension of SST \cite{bib:msst} does not have mode separation capability. It only gives a graphical representation of T-F spectrum thereby limiting its scope of applications to exploratory data analysis only; iii) multivariate extension of EWT \cite{bib:mewt} is simplistic and relies on predefined boundaries of wavelet filterbank which may be difficult to define in real applications. 

In the proposed variational model of MVMD, we seek a collection of common multivariate modulated oscillations residing in input data that exhibit minimum collective bandwidth while fully reconstructing all channels of input data. Multivariate oscillations are modeled using canonical analytic signal representation based on Hilbert transform with a constraint that a joint oscillation (or joint frequency component) exists among all data channels, i.e., multivariate instantaneous frequency. The proposed multivariate variational model is then constructed as a generic extension of standard univariate VMD model to multivariate data. That leads to our proposed method inheriting a lot of desirable properties of standard VMD. Like standard VMD, we solve the optimization problem directly in the Fourier domain through the application of ADMM optimization approach \cite{bib:admm}. The effectiveness of the proposed method is demonstrated through extensive simulations performed on both test (synthetic) signals and real world data sets. We specifically focus on the mode-alignment property of the proposed method for multivariate data that is critical in many real world applications including fusion \cite{bib:Abdullah} and denoising \cite{bib:memd_den}.                         

The remaining part of the paper is organized as follows: Section II introduces the concept of univariate and multivariate modulated oscillations which are utilized in the construction of VMD and the proposed multivariate VMD models respectively. In Section III, we review the standard VMD algorithm. The proposed multivariate VMD method is illustrated in detail in Section IV. To demonstrate the effectiveness of our method, Section V presents detailed experimental results on both synthetic and real world signals and also includes related discussion. The section is divided into two parts: firstly, we show the ability of the proposed method to extract multivariate modulated oscillations (or common rotational modes) in multivariate data, its ability to align common modes across multiple channels and its noise robustness. Finally, we demonstrate the utility of the proposed method through its decomposition capability on real world data sets including multivariate EEG data and cardiotocographic traces.

\section{Univariate and Multivariate Modulated Oscillations}

In this section, we describe mathematical representations of modulated oscillations in single and multidimensional spaces that will be utilized in the construction of standard single-channel VMD and the proposed multivariate VMD models, respectively. We only illustrate important and relevant concepts here; interested readers are referred to \cite{bib:lilly1, bib:lilly2, bib:lilly3} for a detailed treatment on the topic.     

\subsection{Univariate Oscillations}

We start with a univariate AM-FM signal $u(t)$ which is given by

\begin{equation}
u(t)=a(t)\cos(\phi(t)),
\label{eq:uni_amfm}
\end{equation}

\noindent where $a(t)$ is amplitude function while $\phi(t)$ represents time varying phase function due to frequency modulation. The above representation \eqref{eq:uni_amfm} is not unique since more than one functions for $a(t)$ and $\phi(t)$ may be associated with real valued $u(t)$. 

By using an analytic representation of $u(t)$ by means of Hilbert transform, however, a unique \textit{canonic set} of amplitude and phase functions can be defined. That analytic representation $u_+(t)$ is defined by pairing the signal with its own Hilbert transform to yield the following complex signal 

\begin{equation}
u_+(t)=u(t)+j\mathcal{H}u(t)=a(t)e^{j\phi(t)},
\label{eq:uni_analytic}
\end{equation}    

\noindent where the Hilbert transform of $u(t)$ is given by

\begin{equation}
\mathcal{H}u(t)=\frac{1}{\pi}\int_{-\infty}^{\infty}\frac{u(t^\prime)}{t-t^\prime}dt^\prime,
\label{eq:uni_hilbert}
\end{equation}  
      
\noindent where $\int$ is the Cauchy principal integral value. The Fourier transform of $u_+(t)$ yields a unilateral signal spectrum and is defined as

\begin{equation}
\hat{u}_+(\omega)=\int_{-\infty}^{\infty}u_+(t)e^{-j\omega t}dt=1+\text{sgn}(\omega)\hat{u}(\omega).
\label{eq:uni_spec}
\end{equation}  

The original real-valued signal $u(t)$ can finally be recovered from the corresponding complex-valued analytic signal $u_+(t)$ by taking its real part

 \begin{equation}
 u(t)=\mathcal{R}\{u_+(t)\}.
 \end{equation} 

\subsection{Multivariate Oscillations}
To obtain a convenient mathematical form for multivariate oscillations, we first represent a set of $C$ real valued AM-FM signals, denoted by $\mathlarger{\{}u_i(t)\mathlarger{\}}_{i=1}^C$, in a vector form 

\begin{equation}
\mathbf{u}(t)=\begin{bmatrix}
u_{1}(t) \\
u_{2}(t) \\
\vdots \\
u_{c}(t)
\end{bmatrix}=\begin{bmatrix}
a_{1}(t)\cos(\phi_1(t)) \\
a_{2}(t)\cos(\phi_2(t)) \\
\vdots \\
a_{c}(t)\cos(\phi_c(t))
\end{bmatrix}.
\end{equation} 

\noindent where $a_i(t)$ and $\phi_i(t)$ denote amplitude and phase function corresponding to the $i$th signal component respectively.

Next, we obtain an analytic representation of the vector signal $\mathbf{u}(t)$ by employing the Hilbert transform operator, as defined in \eqref{eq:uni_hilbert}, to each component of $\mathbf{u}(t)$ separately. That results in the following analytic signal vector

\begin{eqnarray}
\label{eq:mmod}
\mathbf{u}_+(t)&=&\mathbf{u}(t)+j\mathcal{H}\mathbf{u}(t),\\
&=&\begin{bmatrix}
u_+^{1}(t) \\
u_+^{2}(t) \\
\vdots \\
u_+^{C}(t)
\end{bmatrix}=\begin{bmatrix}
a_{1}(t)e^{j\phi_1(t)} \\
a_{2}(t)e^{j\phi_2(t)} \\
\vdots \\
a_{C}(t)e^{j\phi_C(t)}
\end{bmatrix},
\end{eqnarray}  

\noindent where the canonical (and unique) set of amplitude and phase functions for the $i$th component are given by $a_i(t)=|u_+^{i}(t)|$ and $\phi_i(t)=\text{arg}(u_+^{i}(t))$. Here, ``arg'' denotes the complex argument or the phase function. Finally, the original real-valued signal vector can be obtained as the real part $\mathcal{R}$ of $\mathbf{u}_+(t)$ i.e.,

\begin{equation}
\mathbf{u}(t)=\mathcal{R}\{\mathbf{u}_+(t)\}.
\end{equation}

Note that the above mathematical description of $\mathbf{u}(t)$ considers the $C$ number of AM-FM components in isolation from each other. For a multivariate modulated oscillation, there may be one or more common frequency components present in $\mathbf{u}(t)$. In this work, we consider a simplified multivariate analytic signal model \cite{bib:lilly3} for $\mathbf{u}_+(t)$ that assumes a single common component, i.e., $\omega=\frac{d\phi(t)}{dt}$, among all data channels  

\begin{equation}
\mathbf{u}_+(t)=||\mathbf{u}_+(t)||e^{j\phi(t)}.
\label{eq:mult_osc}
\end{equation}   

The above model will be used in the formulation of the proposed variational decomposition model for multivariate data in Section IV. 

\section{Variational Mode Decomposition}

In this section, we describe the variational mode decomposition (VMD) algorithm for univariate data \cite{bib:vmd}. The goal of VMD is to decompose an input data into a finite and predefined $K$ number of intrinsic principal modes $u_k(t)$, that are similar in form to \eqref{eq:am-fm}, i.e.,

\begin{equation}
x(t)=\sum_{k=1}^{K}u_{k}(t).
\label{eq:vmd}
\end{equation}  

The modes are chosen based on a variational model that aims to minimize the sum of bandwidths of all modes $\mathlarger{\{}u_k(t)\mathlarger{\}}_{k=1}^K$ while also reconstructing the input signal fully or in least squares sense through the addition of modes. In VMD, the bandwidth of each mode $u_k(t)$ is estimated by: i) computing the analytic signal representation $u_+^k(t)$ that exhibits unilateral frequency spectrum; ii) shifting the resulting unilateral spectrum to baseband via harmonic mixing with a complex exponential of frequency $\omega_k$; iii) taking the squared $L^2$ norm of the gradient of the harmonically mixed signal obtained in the previous step. The associated constrained variational optimization problem therefore becomes                        

\begin{equation}
\begin{aligned}
	& \underset{\{u_k\},\{\omega_{k}\}}{\text{minimize}}
	& &\Bigg\{\mathlarger{\sum}_k\Bigg\Vert{\partial_t\Big[u_+^k(t)e^{-j\omega_kt}\Big]\Bigg\Vert^2_2}\Bigg\}\\[10pt]
	& \text{subject to}
	& & \sum_{k}u_k(t) = x(t),
\label{eq:vmd_opt}
\end{aligned}
\end{equation}
	
\noindent where $u_+^{k}(t)$ denotes analytic signal corresponding to $u_k(t)$; the symbol $\partial_t$ represents partial derivative operation with respect to time; $\{u_k\}$ and $\{\omega_{k}\}$ respectively denote sets of all $K$ number of modes and their center frequencies.

Next, an augmented Lagrangian is constructed that renders the problem given in \eqref{eq:vmd_opt} unconstrained. Two penalty terms are added in the Lagrangian: a quadratic term to enforce reconstruction fidelity and a term with Lagrangian multipliers $\lambda$ to ensure that the constraints are fulfilled strictly. The resulting augmented Lagrangian function is given below

\begin{equation}
\begin{aligned}
\mathcal{L}\left({\{u_k\},\{\omega_{k}\},\lambda}\right) = \alpha\ \mathlarger{\sum}_k\Bigg\Vert\partial_t\Big[u_+^k(t) e^{-j\omega_kt}\Big]\Bigg\Vert^2_2\\[10pt]
+\Bigg\Vert x(t)-\sum_{k}u_k(t)\Bigg\Vert_2^2+ \Bigg\langle\lambda(t),x(t)-\sum_{k}u_k(t)\Bigg\rangle.
\label{eq:vmd_lang}
\end{aligned}
\end{equation}

The solution to the original optimization problem as given in \eqref{eq:vmd_opt} can now be found as the saddle point of the Lagrangian function \eqref{eq:vmd_lang}. In VMD, \eqref{eq:vmd_lang} is solved using the alternate direction method of multipliers (ADMM) approach. Using ADMM, the complete optimization problem is solved as a sequence of iterative sub-optimization problems. These sub-problems are convenient to handle since they seek to minimize cost function iteratively for a single parameter/function of interest rather than optimizing cost function for all optimization variables simultaneously. For instance, to update the mode $u_k(t)$, the following sub-optimization problem is considered at the $n$th iteration 

\begin{equation}
\begin{aligned}
u_k^{n+1}\gets\underset{u_k}{\text{arg min }} \mathcal{L}\left(\mathlarger{\{}u_{i<k}^{n+1}\mathlarger{\}},\mathlarger{\{}u_{i\geq k}^n\mathlarger{\}},\mathlarger{\{}\omega_{i}^n\mathlarger{\}},\lambda^n\right). 
\label{eq:vmd_mode1}
\end{aligned}
\end{equation}

The above minimization problem is solved in the spectral domain within VMD $\big($see (19) - (22) in \cite{bib:vmd}$\big)$, resulting in the following update in the frequency domain

\begin{equation}
\begin{aligned}
\hat{u}_k^{n+1}(\omega) = \frac{\hat{x}(\omega)-\sum_{i\neq k}\hat{u}_i(\omega)+\frac{\hat{\lambda}(\omega)}{2}}{1+2\alpha(\omega-\omega_k)^2}.
\end{aligned}
\label{eq:vmd_modeupdate}
\end{equation}

In the next stage, in order to get an update on center frequency $\omega_k$, the following sub-optimization problem is solved iteratively
       	       
\begin{equation}
\begin{aligned}
\omega_k^{n+1}\gets\underset{\omega_k}{\text{arg min }} \mathcal{L}\left({\{u_i^{n+1}\},\{\omega_{i<k}^{n+1}\}\{\omega_{i\geq k}^n\},\lambda^n}\right), 
\end{aligned}
\end{equation}

\noindent to get an update for $\omega_k^{n+1}$ in the dual frequency domain 

\begin{equation}
\omega_k^{n+1}=\frac{\mathlarger\int^{\infty}_0 \omega|\hat{u}_k(\omega)|^2 d\omega}{ \mathlarger\int^{\infty}_0 |\hat{u}_k(\omega)|^2 d\omega},
\label{eq:vmd_frequpdate}
\end{equation}

\noindent which estimates the new frequency as the center of gravity of the associated modes' power spectrum. 

The mode and the center frequency update relations in the spectral domain, given in  \eqref{eq:vmd_modeupdate} and \eqref{eq:vmd_frequpdate} respectively, form the crux of the VMD algorithm. We refer readers to Algorithm 2 in \cite{bib:vmd} for details.

\section{Multivariate Variational Mode Decomposition}

A novel multivariate extension of VMD, namely multivariate VMD (MVMD) is presented in this section. As a generalized extension of the original VMD algorithm for multivariate data residing in multidimensional spaces, the main goal of MVMD is to extract predefined $K$ number of multivariate modulated oscillations $\mathbf{u}_{k}(t)$ from input data $\mathbf{x}(t)$ containing $C$ number of data channels, i.e., $\mathbf{x}(t)=[x_1(t), x_2(t), \cdots x_C(t)]$

\begin{equation}
\mathbf{x}(t)=\sum_{k=1}^{K}\mathbf{u}_{k}(t),
\label{eq:mvmd}
\end{equation}   

\noindent where $\mathbf{u}_{k}(t)=[u_1(t), u_2(t), \cdots u_C(t)]$. 

The goal is to extract an ensemble of multivariate modulated oscillations $\{\mathbf{u}_k(t)\}_{k=1}^K$ in input data such that: i) the sum of bandwidths of the extracted modes is minimum and ii) the sum of the extracted modes exactly recover the original signal $\mathbf{u}_k(t)$. To accomplish that, let us denote the vector analytic representation of  $\mathbf{u}_k(t)$ by $\mathbf{u}_+^k(t)$ (see \eqref{eq:mmod}). The bandwidth of $\mathbf{u}_k(t)$ can then be estimated by taking the $L_2$ norm of the gradient function of the harmonically shifted $\mathbf{u}_+^k(t)$. The resulting cost function  $f$ for MVMD then becomes a multivariate extension of the cost function used in the corresponding VMD optimization problem in \eqref{eq:vmd_opt} and is given by

\begin{equation}
f=\mathlarger{\sum}_k\Bigg\Vert\partial_t\Big[e^{-j\omega_kt}\mathbf{u}_+^k(t)\Big]\Bigg\Vert^2_2
\label{eq:mvmd_cost1}
\end{equation}

Another important point to highlight in \eqref{eq:mvmd_cost1} is that a single frequency component $\omega_k$ is used in the harmonic mixing of the whole vector $\mathbf{u}_+^{k}(t)$. By design, therefore, we are looking to find multivariate oscillations in $\mathbf{u}_k(t)$ that have a single common frequency component $\omega_k$ in all channels. That is exactly how we defined our model for multivariate modulated oscillation in   
\eqref{eq:mult_osc}. The bandwidth of modulated multivariate oscillation is therefore estimated by shifting the unilateral frequency spectrum of \textit{all channels} of $\mathbf{u}_+^{k}(t)$ by $\omega_k$ and taking the Frobenius norm \footnote{The Frobenius norm is taken as a direct extension of the $L_2$ norm used in original VMD to the matrices which appear due to multiple channel signal representation in MVMD.} of the resulting matrix. That leads to the following convenient representation of $f$

\begin{equation}
f=\mathlarger{\sum_{k}\sum_c}\Bigg\Vert{\partial_t\Big[u_+^{k,c}(t) e^{-j\omega_kt}\Big]\Bigg\Vert^2_2}
\label{eq:mvmd_cost2}
\end{equation}

\noindent where $u_+^{k,c}(t)$ denotes the analytic modulated signal corresponding to channel number $c$ and mode number $k$. In contrast to the vector signal in \eqref{eq:mvmd_cost1}, $u_+^{k,c}(t)$ is a complex-valued signal with single component in \eqref{eq:mvmd_cost2}. We now present the constrained optimization problem for MVMD   

\begin{equation}
\begin{aligned}
& \underset{\{u_{k,c}\},\{\omega_{k}\}}{\text{minimize}}
& & \Bigg\{ \mathlarger{\sum_{k}\sum_c}\Bigg\Vert\partial_t\Big[u_+^{k,c}(t) e^{-j\omega_kt}\Big]\Bigg\Vert^2_2\Bigg\}\\[10pt]
& \text{subject to}
& & \sum_{k}u_{k,c}(t) = x_c(t),\text{   } c=1,2, \ldots ,C.
\label{eq:mvmd_opt}
\end{aligned}
\end{equation}
   
Note that there are multiple linear equality constraints in the above model corresponding to the total number of channels. The corresponding augmented Lagrangian function then becomes

\begin{equation}
\begin{aligned}
\mathcal{L}\left({\{u_{k,c}\},\{\omega_{k}\},\lambda_c}\right) = \alpha\ \mathlarger{\sum_k\sum_c}\Bigg\Vert\partial_t\Big[u_+^{k,c}(t) e^{-j\omega_kt}\Big]\Bigg\Vert^2_2\\[10pt]
+\mathlarger{\sum_c}\Bigg\Vert x_c(t)-\sum_{k}u_{k,c}(t)\Bigg\Vert_2^2+ \mathlarger{\sum_c}\Bigg\langle\lambda_c(t),x_c(t)-\sum_{k}u_{k,c}(t)\Bigg\rangle.
\label{eq:mvmd_lang}
\end{aligned}
\end{equation} 

The above unconstrained optimization problem is solved using ADMM approach as illustrated in Algorithm 1. Note from Algorithm 1 that how the ADMM approach converts a complex optimization problem \eqref{eq:mvmd_opt} into multiple more simpler suboptimization problems, as given by the update equations (25)-(27).  

\begin{algorithm}[t]
	\caption{\bf Illustration of ADMM for MVMD Optimization}
	\label{alg:mvmd_time}
	\vspace{2mm}\text{Initialize: } \vspace{-4.5mm}\begin{equation*}\{u_{k,c}^1\}\mbox{, }\{\omega_k^1\}\mbox{, }\lambda^1\mbox{, }n\gets 0\end{equation*}
	\vspace{-3mm}
	\begin{algorithmic} 
		\Repeat
		\State $n \gets n+1$
		\vspace{.2cm}
		\For{$k=1:K$}
		\vspace{.1cm}
		\For{$c=1:C$}
		\vspace{.1cm}
		\textit{Update mode $u_{k,c}$:}
		\vspace{.1cm}
		\State  \begin{equation} u_{k,c}^{n+1}\gets\underset{u_{k,c}}{\text{arg min }} \mathcal{L}\left(\mathlarger{\{}u_{i<k,c}^{n+1}\mathlarger{\}},\mathlarger{\{}u_{i\geq k,c}^n\mathlarger{\}},\mathlarger{\{}\omega_{i}^n\mathlarger{\}},\lambda_c^n\right) \label{eq:mvmd_uk}  \end{equation}
		\EndFor
		\vspace{.1cm}
		\EndFor
		\vspace{.2cm}
		\For{$k=1:K$}
		\vspace{.1cm}
		\textit{Update center frequency $\omega_{k}$:}
		\State \begin{equation}\omega_k^{n+1}\gets\underset{\omega_{k}}{\text{arg min }} \mathcal{L}\left({\{u_{i,c}^{n+1}\},\{\omega_{i< k}^{n+1}\}\{\omega_{i\geq k}^n\},\lambda_c^n}\right) \label{eq:mvmd_wk} \end{equation} 
		\EndFor
		\vspace{.1cm}
		\vspace{.1cm}
		\For{$c=1:C$}
		\vspace{.1cm}
		\textit{Update $\lambda_c$:}
		\State \begin{equation}\lambda_c^{n+1}=\lambda_c^n+\tau \Big(x_c-\sum_k u_{k,c}^{n+1}\Big) \label{eq:mvmd_lambdatime} \end{equation} 	
		\EndFor
		\vspace{.1cm}
		\Until{Convergence: $\sum_k\sum_c \frac{\Vert u_{k,c}^{n+1}-u_{k,c}^n\Vert_2^2}{\Vert u_{k,c}^n\Vert_2^2}<\epsilon$} 
\end{algorithmic}
\end{algorithm}

\subsection{Mode update}
We first focus on the minimization problem related to the mode update \eqref{eq:mvmd_uk}. Its equivalent optimization problem is given below 
        
\begin{equation}
\begin{aligned}
	u_{k,c}^{n+1}=\underset{u_{k,c}}{\text{arg min}} \Bigg\{\alpha\Bigg\Vert\partial_t\Big[u_+^{k,c}(t) e^{-j\omega_kt}\Big]\Bigg\Vert^2_2+\\
	\Bigg\Vert x_c(t)-\mathlarger{\sum_i} u_{i,c}(t)+\frac{\lambda_c(t)}{2}\Bigg\Vert^2_2\Bigg\}
\end{aligned}
\end{equation}

This subproblem is similar in form to the mode update problem of original VMD \eqref{eq:vmd_mode1}. There, the problem was solved in the Fourier domain to yield a convenient update relation in the frequency domain \eqref{eq:vmd_modeupdate}. We use that result to give the following mode update relation in our case

\begin{equation}
\begin{aligned}
\hat{u}_{k,c}^{n+1}(\omega) = \frac{\hat{x_c}(\omega)-\sum_{i\neq k}\hat{u}_{i,c}(\omega)+\frac{\hat{\lambda}_c(\omega)}{2}}{1+2\alpha(\omega-\omega_k)^2}.
\end{aligned}
\label{eq:mvmd_modeupdate}
\end{equation}

\subsection{Center frequency update}
Now we turn our attention to the optimization problem corresponding to center frequency update \eqref{eq:mvmd_wk}. Considering that the last two terms of the Lagrangian \eqref{eq:mvmd_lang} do not depend on $\omega_k$, the relevant problem simplifies to

\begin{equation}
\omega_{k}^{n+1}=\underset{\omega_k}{\text{arg min}}\Bigg\{\mathlarger{\sum_c}\Bigg\Vert\partial_t\Big[u_+^{k,c}(t) e^{-j\omega_kt}\Big]\Bigg\Vert^2_2\Bigg\}.
\label{eq:mvmd_wkupdate}
\end{equation}
 
The optimization can be performed in the frequency domain more conveniently. Using the Plancherel theorem that relates the inner product of a function in time and frequency domain, we obtain an equivalent problem of \eqref{eq:mvmd_wkupdate} in the Fourier domain which is given below 

\begin{equation}
\omega_k^{n+1}=\underset{\omega_k}{\text{arg min}}\Bigg\{\mathlarger{\sum_c}\int_{0}^{\infty}(\omega-\omega_k)^2\Big| \hat{u}_{k,c}(\omega)\Big|^2d\omega\Bigg\}.
\end{equation}

Minimizing the above sum of quadratic functions by setting its first derivative to zero, followed by a few simple algebraic manipulations yields the following relation

\begin{equation}
\omega_k^{n+1}=\frac{\mathlarger{\sum_c}\mathlarger\int^{\infty}_0 \omega|\hat{u}_{k,c}(\omega)|^2 d\omega}{\mathlarger{\sum_c}\mathlarger\int^{\infty}_0 |\hat{u}_{k,c}(\omega)|^2 d\omega}.
\label{eq:mvmd_frequpdate}
\end{equation}

By comparing the above frequency update relation of MVMD with the corresponding relation for VMD \eqref{eq:vmd_frequpdate}, we note that while updating $\omega_k$ for each mode in the proposed MVMD method, contributions from the power spectrum of all $C$ number of channels is taken into account. 

Having the mode \eqref{eq:mvmd_modeupdate} and the center frequency update relations \eqref{eq:mvmd_frequpdate} in the frequency domain at our disposal, we can perform the optimization in frequency domain using the ADMM approach; the steps are listed in Algorithm 2.

\begin{algorithm}[H]
	\caption{\bf Multivariate VMD}
	\label{alg:mvmd_freq}
		\vspace{2mm}\text{Initialize: } \vspace{-4.5mm}\begin{equation*}\{\hat{u}_{k,c}^1\}\mbox{, } \{\omega_k^1\}\mbox{, } \hat{\lambda}_c^1\mbox{, }n\gets 0\end{equation*}
		\vspace{-3mm}
	\begin{algorithmic} 
		\Repeat 
		\State $n \gets n+1$
		\For{$k=1:K$}
		\vspace{.1cm}
		\For{$c=1:C$}
		\textit{Update mode $\hat{u}_{k,c}$:}
		\State  
		\begin{equation}  
		\hat{u}_{k,c}^{n+1}(\omega) \gets \frac{\hat{x}_c(\omega)-\sum_{i\neq k}\hat{u}_{i,c}(\omega)+\frac{\hat{\lambda}_c^{n}(\omega)}{2}}{1+2\alpha(\omega-\omega_k^n)^2}
		\label{eq:mvmd_hatuk}  
		\end{equation} 
		\EndFor
		\vspace{.1cm}
		\EndFor
		\vspace{.2cm}
		\For{$k=1:K$}
		\textit{Update center frequency $\omega_{k}$:}
		\State 
		\begin{equation} 
		\omega_k^{n+1}\gets \frac{\mathlarger{\sum_c}\mathlarger\int^{\infty}_0 \omega|\hat{u}_{k,c}^{n+1}(\omega)|^2 d\omega}{\mathlarger{\sum_c}\mathlarger\int^{\infty}_0 |\hat{u}_{k,c}^{n+1}(\omega)|^2 d\omega}
		\label{eq:mvmd_hatwk} 
		\end{equation} 
		\EndFor
		\vspace{.1cm}
		\For{$c=1:C$}
		\textit{Update $\lambda_c$}:\\
		\mbox{   }for all $\omega \geq 0$
		\State \begin{equation}\hat{\lambda}_c^{n+1}(\omega)=\hat{\lambda}_c^n(\omega)+\tau \Big(\hat{x}_c(\omega)-\sum_k \hat{u}_{k,c}^{n+1}(\omega)\Big) \label{eq:mvmd_lambda} \end{equation} 	
		\EndFor
		\Until{Convergence: $\sum_k\sum_c \frac{\Vert \hat{u}_{k,c}^{n+1}-\hat{u}_{k,c}^n\Vert_2^2}{\Vert \hat{u}_{k,c}^n\Vert_2^2}<\epsilon$} 
\end{algorithmic}
\end{algorithm}

\begin{figure}[!t]
	\centering
	\begin{subfigure}{0.5\textwidth}
		\centering
		\includegraphics[trim={5mm 0mm 5mm 5mm},clip,width=1\linewidth,height=0.65\linewidth]{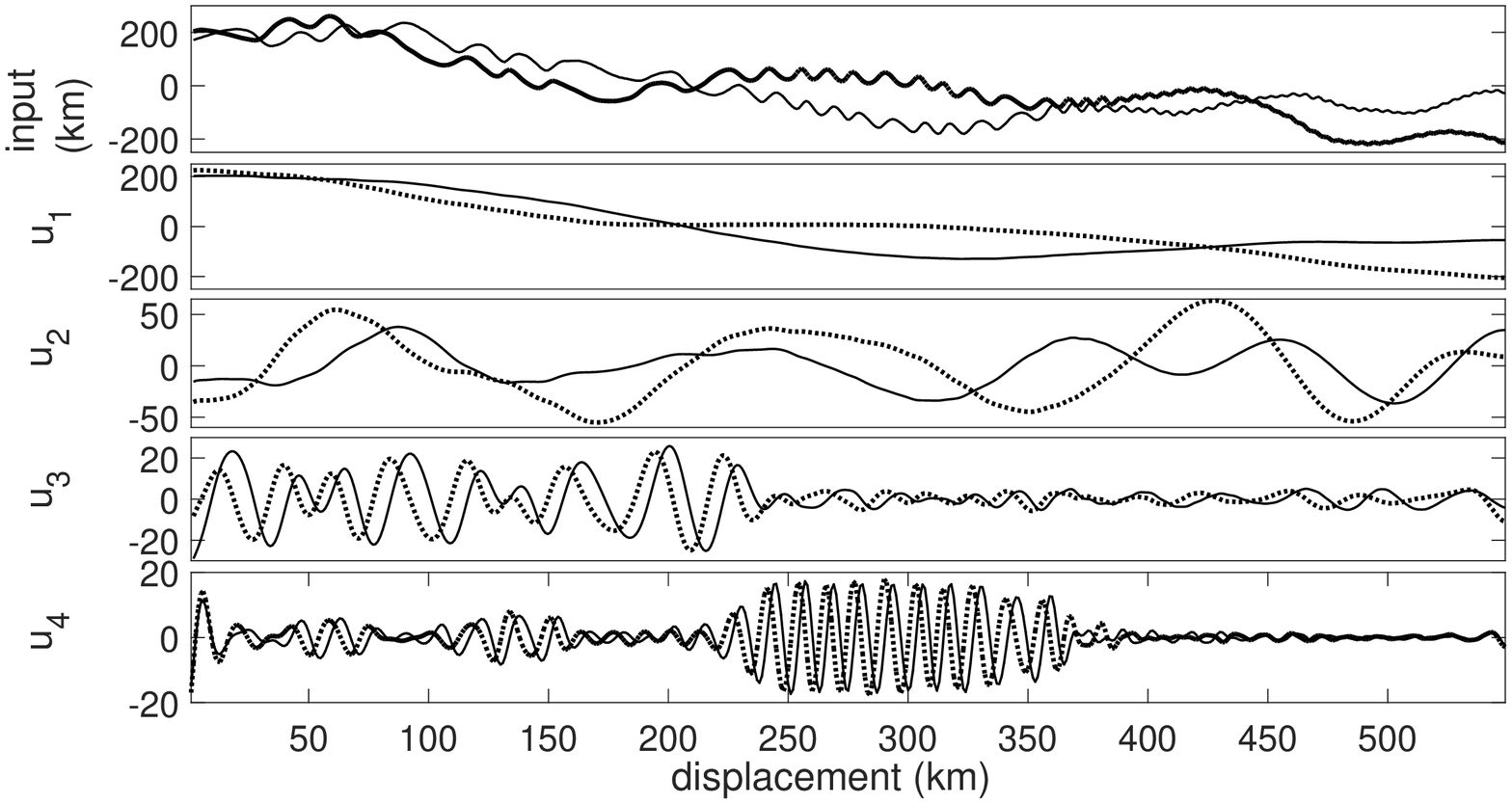}
		\centering
		\caption{Time plots of bivariate modulated oscillations}
	\end{subfigure}
	\begin{subfigure}{0.5\textwidth}
		\centering
		\includegraphics[trim={5mm 0mm 0mm 0mm},clip,width=1\linewidth]{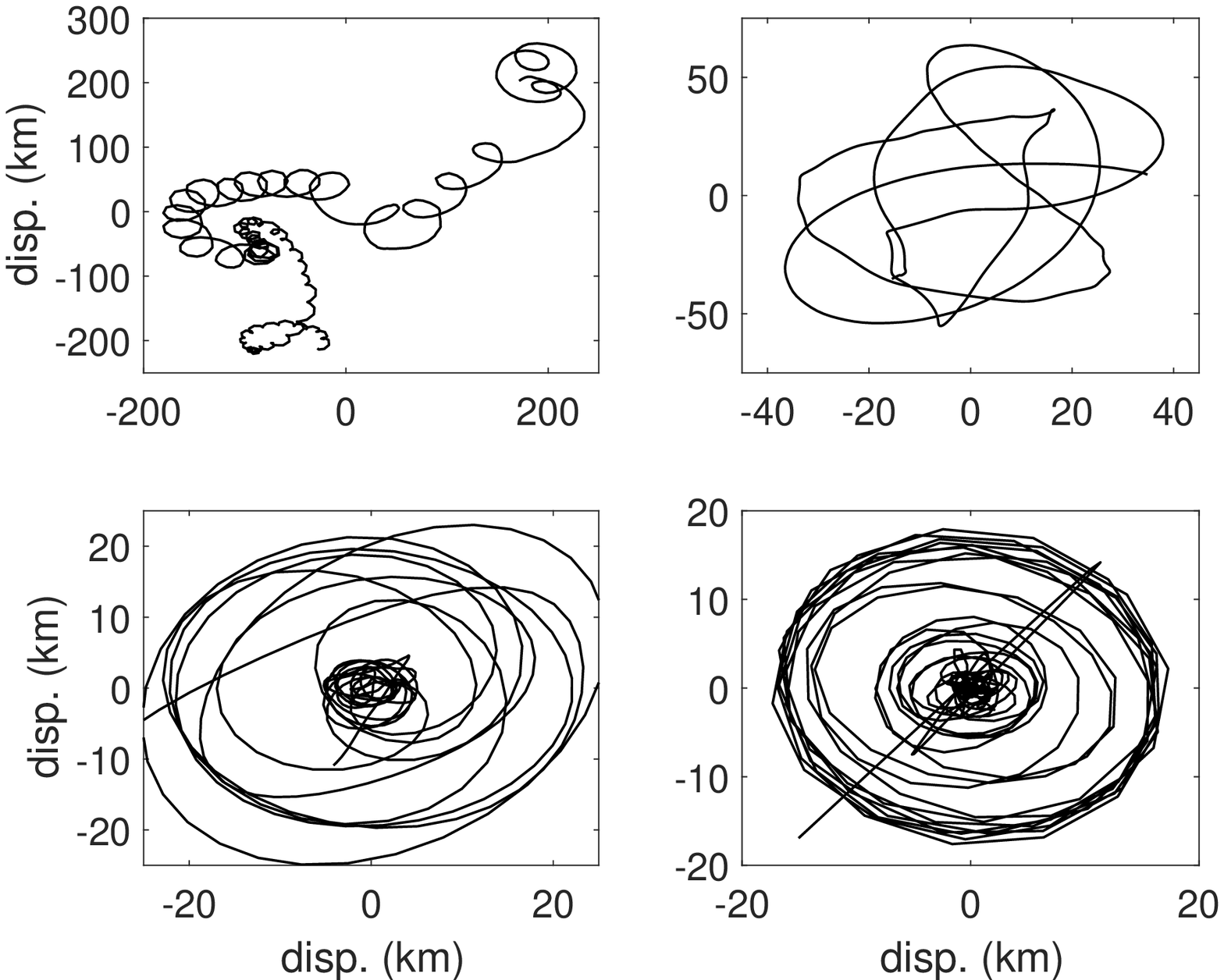}
		\caption{Bivariate modulated oscillations in 2D}
	\end{subfigure}
	\caption{Illustration of bivariate modulated oscillations obtained from applying MVMD algorithm to position record of an oceanographic float. a) Time plots of the two-channel MVMD modes; b) Modulated oscillations in 2D corresponding to original signal (top left) and mode 2 (top right), 3 (lower left) and 4 (lower right).}
	\label{fig:rot}
\end{figure}

\section{Experimental Results}
To evaluate the performance of the proposed method, we conducted experiments and simulations on a wide range of multivariate signals; their detailed results are presented in this section. 

Being a direct extension of the original VMD algorithm for multivariate data, MVMD inherits all the important advantages of the original VMD such as its robustness to noise and low sampling rates and the ability to separate tones very effectively. All these properties of VMD have been explored in detail in \cite{bib:vmd} and we will not verify them for MVMD, in this section. Instead, we will focus on the ability of the proposed method to align common frequency scales across multiple data channels. That is a critical requirement in many scientific and engineering applications involving multivariate data such as image fusion \cite{bib:memd_fus}, biomedical signal based classification \cite{bib:memd_seizure} and denoising \cite{bib:memd_den}, to name a few.    

We first show the ability of the proposed method to decompose a real world bivariate data into its inherent bivariate modulated oscillations, which are also referred to as \textit{rotational modes}. Secondly, we demonstrate the mode-alignment property of MVMD on synthetic test signals containing a combination of tones across its different channels and white Gaussian noise (wGn). We also compare the results with those obtained from applying original VMD to each channel separately. We next illustrate the effect of noise in input data on mode-alignment property of MVMD. We show comparison of our method against multivariate EMD (MEMD) algorithm. Finally, the effectiveness of the proposed method is shown on a couple of real world data sets including multivariate EEG and bivariate cardiotocographic (CTG) data.              

\begin{figure}[!t]
	\begin{subfigure}{0.5\textwidth}
		\includegraphics[trim={12mm 0 12mm 0},clip,width=1\linewidth]{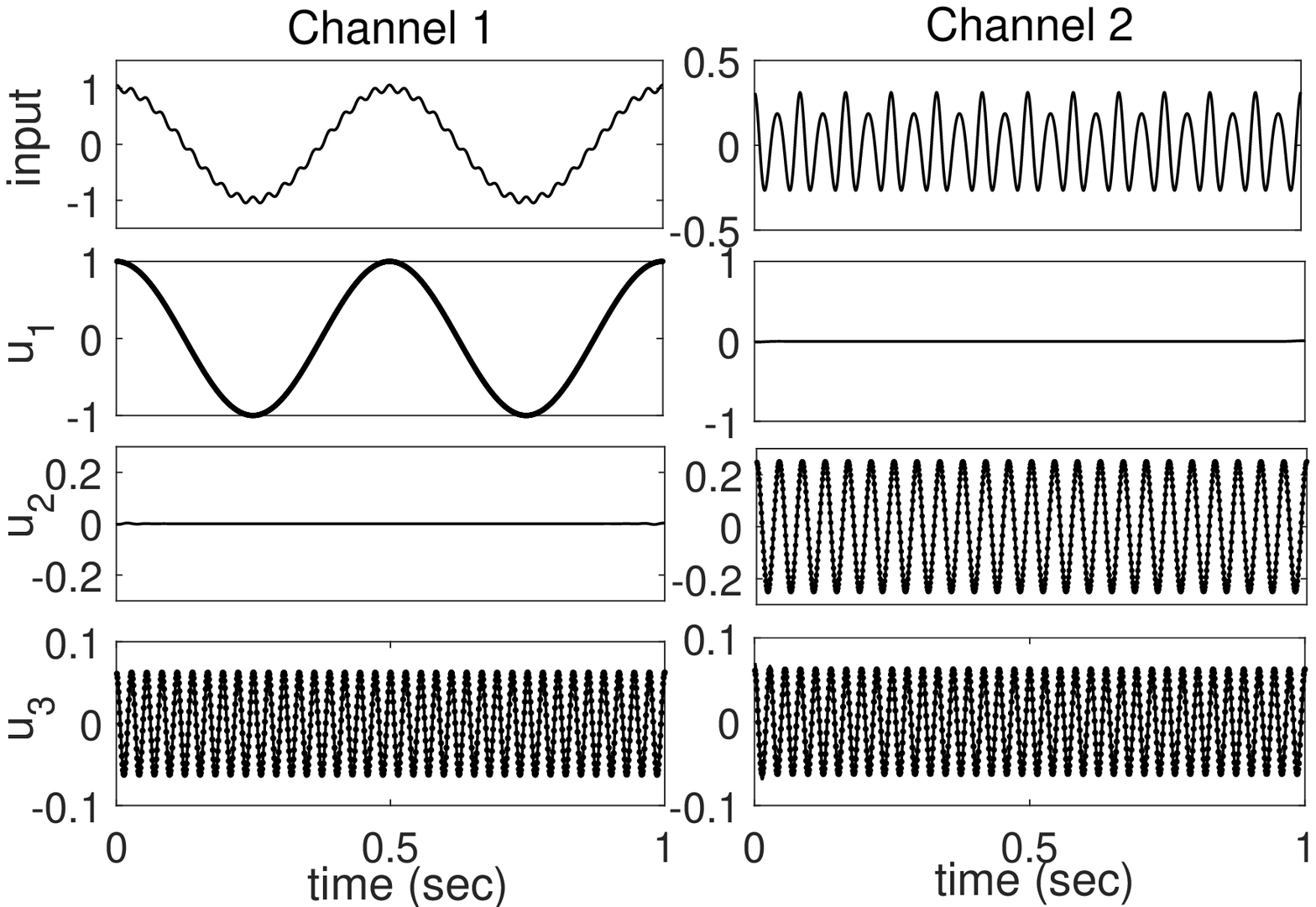}
		\caption{Illustration of MVMD mode-alignment}
	\end{subfigure}
	\begin{subfigure}{0.5\textwidth}
		\includegraphics[trim={12mm 0mm 12mm 0mm},clip,width=0.985\linewidth]{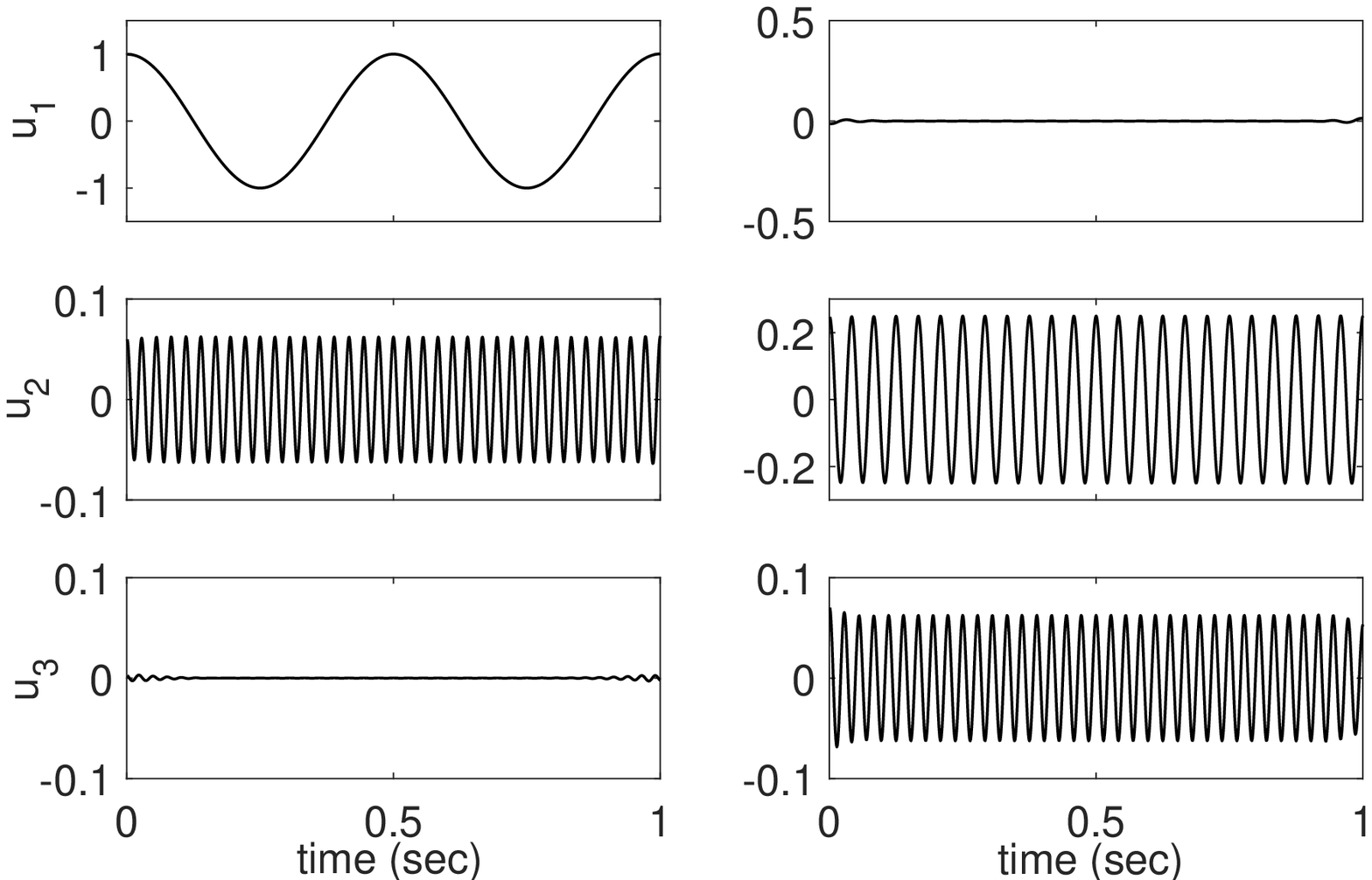}
		\caption{Mode misalignment within VMD}
	\end{subfigure}
	\caption{Decomposition of a bivariate signal consisting of a mixture of tones via a) MVMD; b) VMD applied channel-wise to the two channels separately. Note how the similar frequency modes are aligned in MVMD decomposition while mode 2 within VMD decomposition is misaligned.}
	\label{fig:mode}
\end{figure}
\subsection{Separation of multivariate modulated oscillations}

We demonstrate the ability of the proposed method to separate multivariate modulated oscillations or rotational modes from a real bivariate data set. The data was taken during the Eastern Basin experiment and consists of position record of a subsurface oceanographic float that was deployed in North Atlantic ocean to track the trajectory of salty water flowing from the Mediterranean Sea. The data can be downloaded from the World Ocean Circulation Experiment Subsurface Float Data Assembly Center (WFDAC) at http://wfdac.whoi.edu.

Separate input channels of the data are shown in Figure \ref{fig:rot}(a) (top row) along with the illustration of the data in 2D space (see Figure \ref{fig:rot}(b) (top left)) to visualize multivariate oscillations. From the graphical 2D representation of the input data, bivariate modulated oscillations (or 2D rotations) are quite evident. Looking at the corresponding time plots, joint or common frequency scales across data channels can be seen that explains the presence of rotations or multivariate modulated oscillations in the 2D plot of the data (See Section II.B for detail).      

We applied the proposed MVMD to the bivariate data with an aim to separate its principal multivariate modulated oscillations. We decomposed the data into $K=4$ modes which are shown in Figure \ref{fig:rot}, both as time plots (Figure \ref{fig:rot}(a)) and 2D graphical plots (Figure \ref{fig:rot}(b)). Note from the 2D plot that the first two modes consist of multivariate modulated oscillations or rotating modes which may correspond to the presumed coherent vortex. The non-rotating modes are characterized by the absence of joint frequency scale in one of the channels. This example demonstrates the ability of MVMD to effectively separate multivariate oscillations from input data while also verifying the mode-alignment property of the algorithm on a practical data set.   

\begin{figure}[!t]
	\begin{subfigure}{0.5\textwidth}
		\includegraphics[trim={12mm 7mm 8mm 0},clip,width=0.9\linewidth,height=0.45\linewidth]{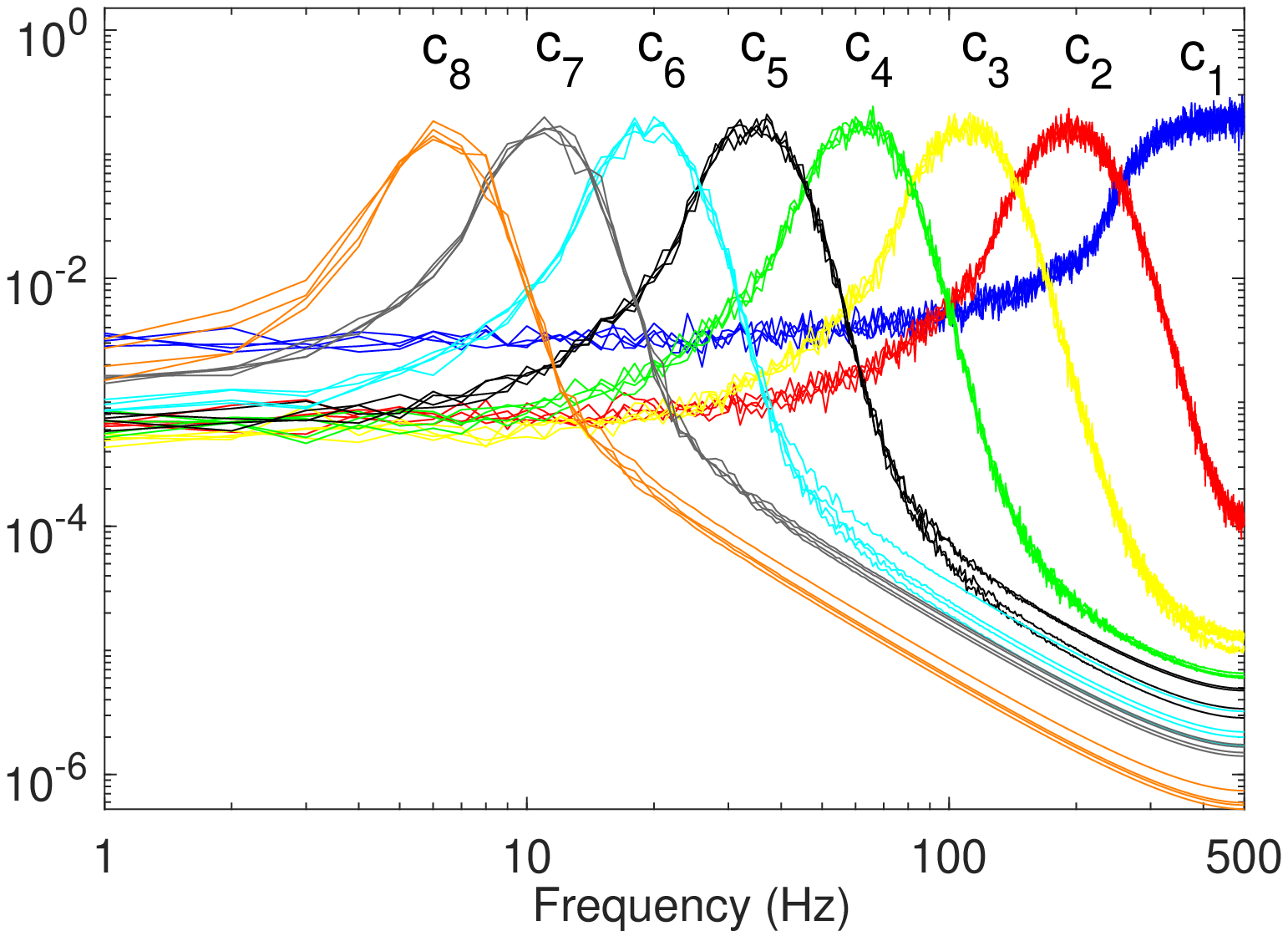}
		\caption{MEMD filterbank for wGn}
	\end{subfigure}
	\begin{subfigure}{0.5\textwidth}
		\includegraphics[trim={12mm 0mm 8mm 0mm},clip,width=0.9\linewidth,height=0.475\linewidth]{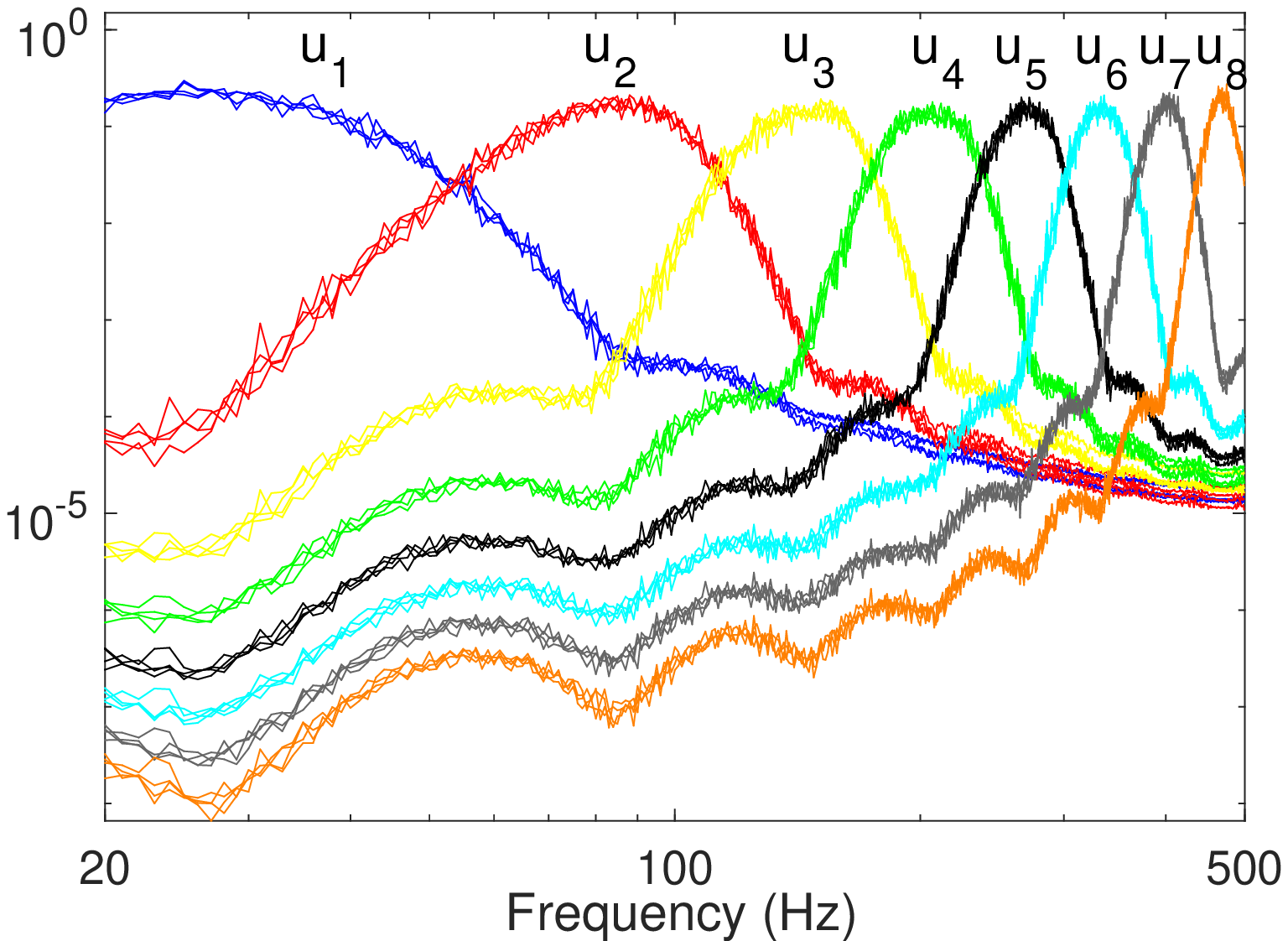}
		\caption{MVMD filterbank for wGn}
	\end{subfigure}
	\caption{Filterbank structure for a) MEMD and b) MVMD for 4-channel wGn. MVMD follows a different filterbank structure as compared to the quasi-dyadic filterbank of MEMD. The mode-alignment across channels is also apparent in both cases.}
	\label{fig:fb}
\end{figure}

\subsection{Mode-alignment property}

Mode-alignment of multivariate data refers to the alignment of common or joint oscillations (containing similar frequency content) across multiple channels of a single mode. This property is of fundamental importance to ensure coherent multivariate T-F analysis in many practical applications involving multivariate data such as fusion, denoising and classification. Typically, in such applications, signal processing methods are applied to each mode separately followed by their reconstruction. In case a mode contains oscillations that exhibit different frequency content across multiple channels, application of signal processing methods will result in physically meaningless estimates or decisions due to misalignment of similar information content across channels. Readers are referred to \cite{bib:looney} for detailed discussion on the importance of mode-alignment in fusion data related application.

We illustrate the ability of MVMD to identify and align principal modulated oscillations present in the data across multiple channels and modes in Figure \ref{fig:mode}(a). The input data was a bivariate signal whose individual components were a mixture of a 36-Hz sinusoid that was common to both the data channels; a 2-Hz tone in the channel-1 and a 24-Hz tone in the channel-2. The $K=3$ number of modes obtained by applying MVMD to the above dataset are shown. Observe that all modes are aligned in terms of their frequency content: the 36-Hz tone present in all data channels is localized in a single mode $u_3$. The 2-Hz signal is located in the channel-1 of $u_1$ while the 24-Hz tone is localized in the channel-2 of $u_2$. 
In Figure \ref{fig:mode}(b), we illustrate that similar mode-alignment across channels is not possible by applying VMD to each channel separately. Notice that frequency content across channels in mode $u_2$ is not aligned i.e., 36-Hz in channel-1 and 24-Hz in channel-2.

\subsection{MVMD vs MEMD filterbanks for wGn}
In \cite{bib:emd_fb, bib:memd_fb}, EMD and MEMD had respectively been shown to exhibit quasi-dyadic filterbank structure for wGn, which is similar to the wavelet filterbank. That has led to diverse range of applications of (M)EMD in science and engineering. We show here that while MVMD also exhibits filterbank structure for wGn, it appears to be different from the quasi-dyadic structure observed within EMD based algorithms. 

To verify that, we applied MVMD on 50 realizations of a four-channel wGn process of length $L = 1000$. We also applied MEMD to the same data set. The power spectral density (PSD) plots averaged over 100 realizations are plotted for both MVMD and MEMD in Figure \ref{fig:fb}. By  comparing the two set of plots, we can observe that MEMD exhibits quasi-dyadic filterbank structure for wGn as evident by almost similar bandwidths of different IMFs in the log-frequency domain. On the other hand, MVMD modes exhibit different bandwidths in the log-frequency domain. A thorough investigation of the MVMD filterbanks is beyond the scope of this paper and would be a good topic for future research.        

Also observe from the Figure \ref{fig:fb} that there is  alignment of frequency content among different channels within similar modes of both MVMD and MEMD: alignment within MVMD appears slightly more prominent as compared to MEMD, resulting in well defined subband filters when compared against MEMD. 

\begin{figure}[!t]

	\begin{subfigure}{0.5\textwidth}
		\centering
		\includegraphics[trim={0mm 0 0mm 0},clip,width=0.75\linewidth,height=0.5\linewidth]{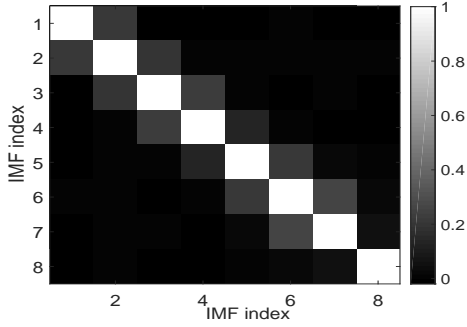}
		\caption{Quasi-orthogonality of modes within MEMD for wGn}
	\end{subfigure}
	\begin{subfigure}{0.5\textwidth}
		\centering
		\includegraphics[trim={0mm 0mm 0mm 0mm},clip,width=0.77\linewidth,height=0.5\linewidth]{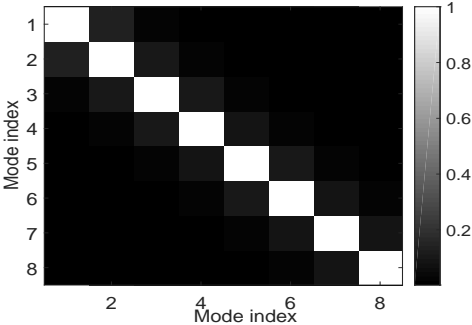}
		\caption{Quasi-orthogonality of modes within MVMD for wGn}
	\end{subfigure}
	\caption{Illustration of quasi-orthogonality of decomposed modes obtained from a) MEMD and b) MVMD for a 4-channel wGn. Cross correlation coefficient matrix \eqref{eq:coerr} is plotted in both cases. Note the almost diagonal and tridiagonal nature of the correlation matrix in MVMD and MEMD respectively highlighting the superiority of MVMD in terms of obtaining quasi-orthogonal modes.}
	\label{fig:lcc}
\end{figure}

\subsection{Quasi-orthogonality of MVMD modes}

In linear signal decomposition and T-F methods such as STFT and wavelet transform, predefined basis functions are by construction chosen to be orthogonal. That ensures that there is no `leakage' of information across different modes (and channels). Since basis functions within data driven decomposition and T-F methods are adaptive in nature, it is important to demonstrate quasi-orthogonality empirically. 

In this section, we establish the quasi-orthogonality of MVMD modes for multiple realizations of the four-channel wGn process, which were used in the previous section to demonstrate the filterbank structure for MVMD and MEMD. Here, we use \textit{correlation coefficient} as a means to quantify the dependence between different modes that were generated from MVMD and MEMD. Specifically, to compute the correlation coefficient between the modes $u_i$ and $u_j$, whose variances are denoted by $\sigma_i$ and $\sigma_j$ respectively, we have

\begin{equation}
\rho_{ij}=\frac{\text{cov}\left(u_i,u_j\right)}{\sigma_i\sigma_j},
\label{eq:coerr}
\end{equation}

\noindent where $\text{cov}(.)$ denotes the covariance between a set of signals.

If the correlation coefficient is close to zero for a set of modes, those modes can be considered to be quasi-orthogonal; a value close to unity suggests very strong correlation. We obtained the \textit{correlation coefficient matrix} for the set of $K=8$ modes obtained from MVMD and MEMD for 4-channel wGn, averaged them over 100 realizations, and plotted the resulting matrix as gray-scale images in Figure \ref{fig:lcc}. Notice the almost diagonal structure of the correlation matrix in the case of MVMD, suggesting a strong quasi-orthogonality among MVMD modes. The correlation matrix corresponding to MEMD, on the other hand, shows some `leakage' between adjacent modes as apparent from its tridiagonal nature.         
     
\begin{figure}[!t]
	\begin{subfigure}{0.5\textwidth}
		\includegraphics[trim={12mm 0 12mm 0},clip,width=1\linewidth]{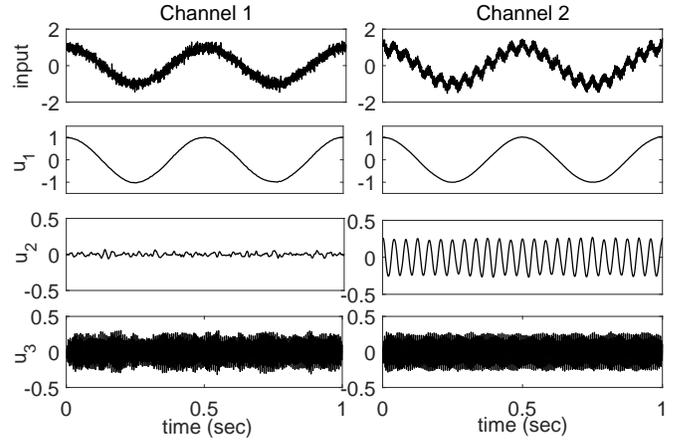}
		\caption{Illustration of robustness of MVMD to noise}
	\end{subfigure}
	\begin{subfigure}{0.5\textwidth}
		\includegraphics[trim={15mm 0mm 18mm 0mm},clip,width=1.02\linewidth]{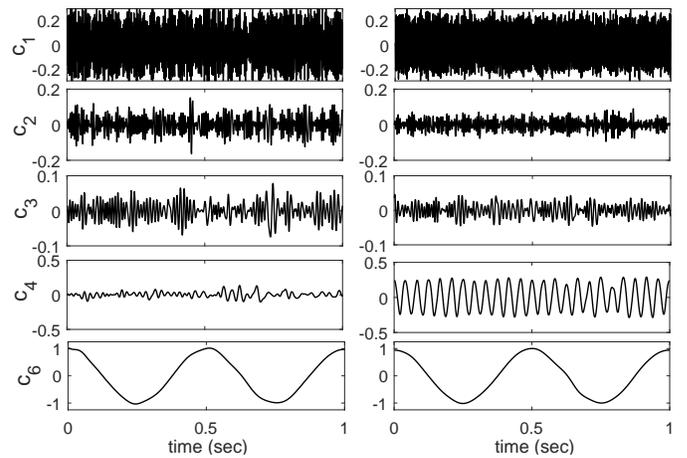}
		\caption{MEMD decomposition}
	\end{subfigure}
	\caption{Decomposition of a bivariate signal consisting of a mixture of tones + wGn in both channels via a) MVMD; b) MEMD. Similar frequency modes are clearly aligned in the MVMD decomposition. In the MEMD decomposition, there is alignment of modes, however, there is some `leakage' of information content in channel 1 of mode $c_4$. Also, 288-Hz tone is spread across the first three IMFs which is undesirable.}
	\label{fig:noisemode}
\end{figure}

\begin{figure*}[!t]
	\centering
	\begin{subfigure}{0.25\textwidth}
		\centering
		\includegraphics[trim={5mm 0mm 8mm 5mm},clip,width=1.0\linewidth]{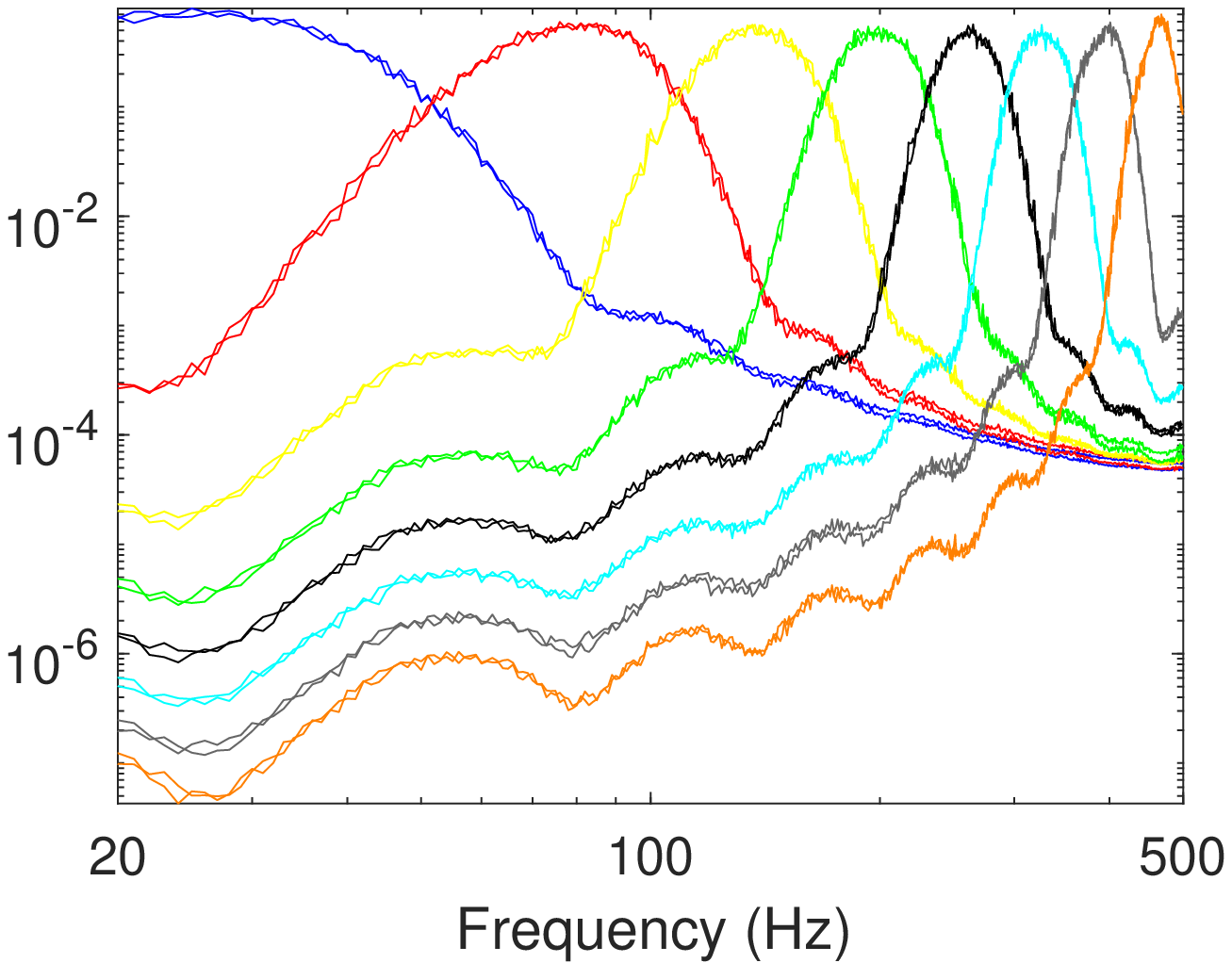}
	\end{subfigure}
	\hspace{-3mm}
	\begin{subfigure}{0.25\textwidth}
		\centering
		\includegraphics[trim={5mm 0mm 8mm 5mm},clip,width=1.0\linewidth]{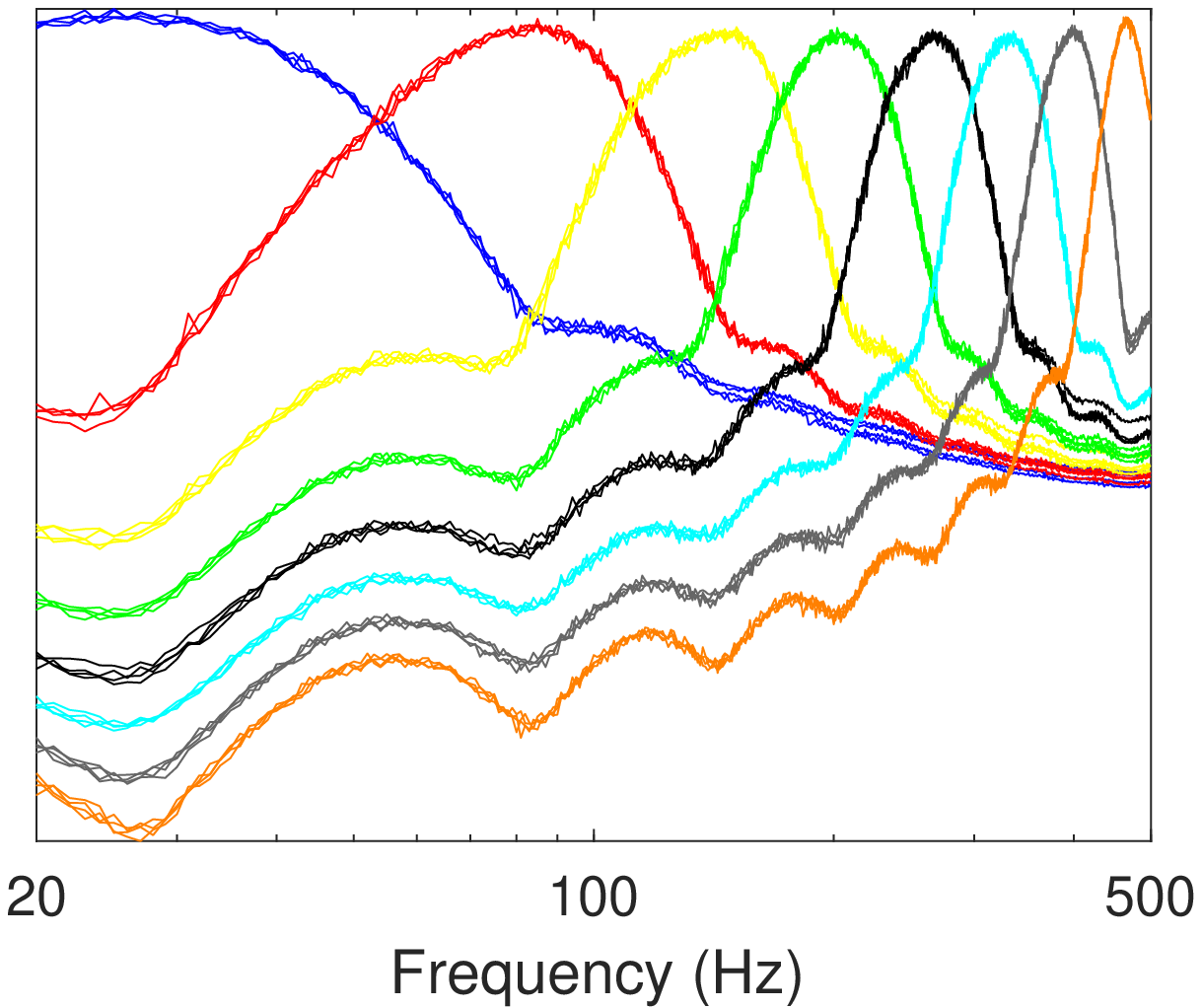}
	\end{subfigure}
	\hspace{-3mm}
	\begin{subfigure}{0.25\textwidth}
		\centering
		\includegraphics[trim={5mm 0mm 8mm 5mm},clip,width=1.0\linewidth]{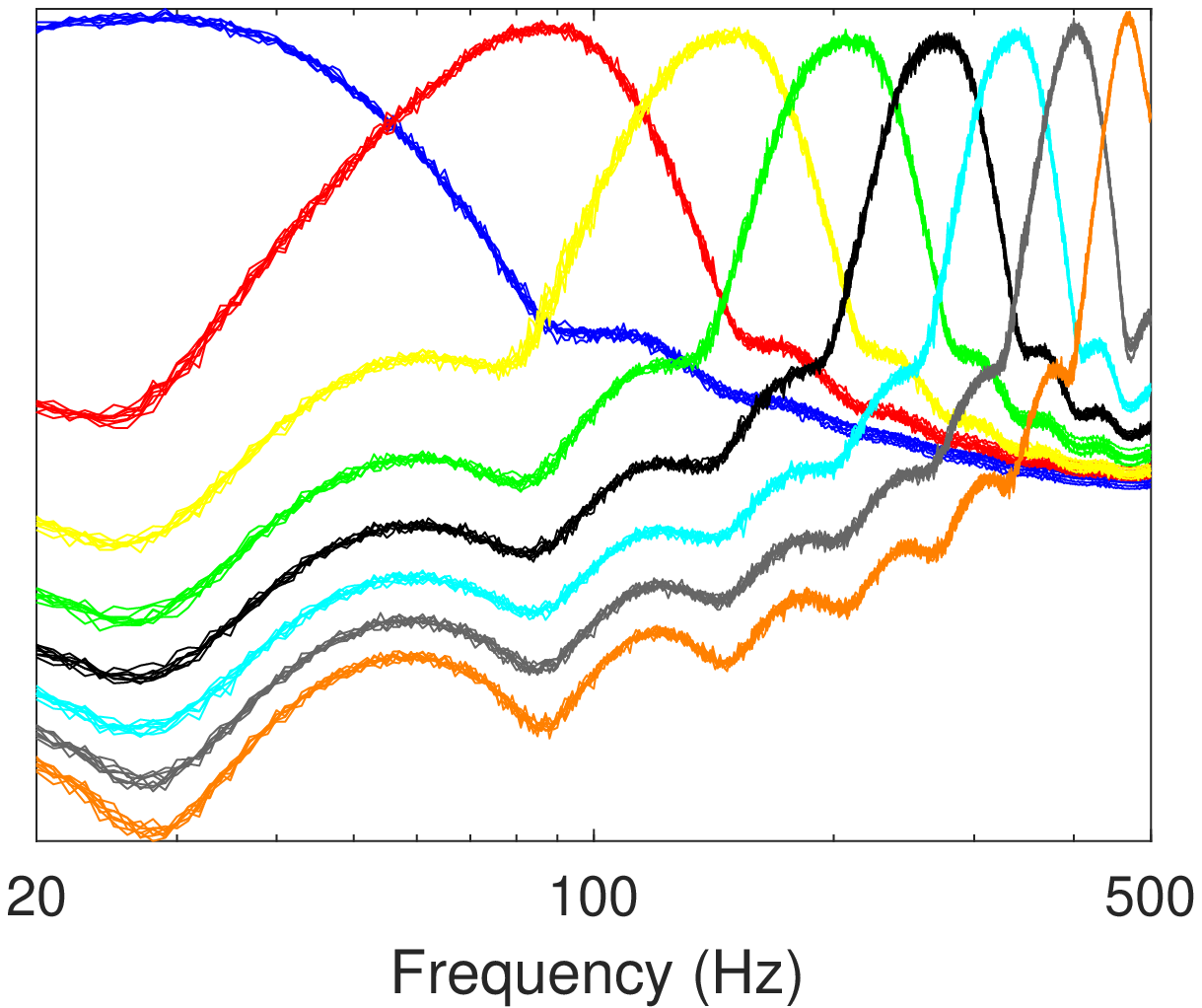}
	\end{subfigure}
	\hspace{-3mm}
	\begin{subfigure}{0.25\textwidth}
		\centering
		\includegraphics[trim={5mm 0mm 8mm 5mm},clip,width=1.0\linewidth]{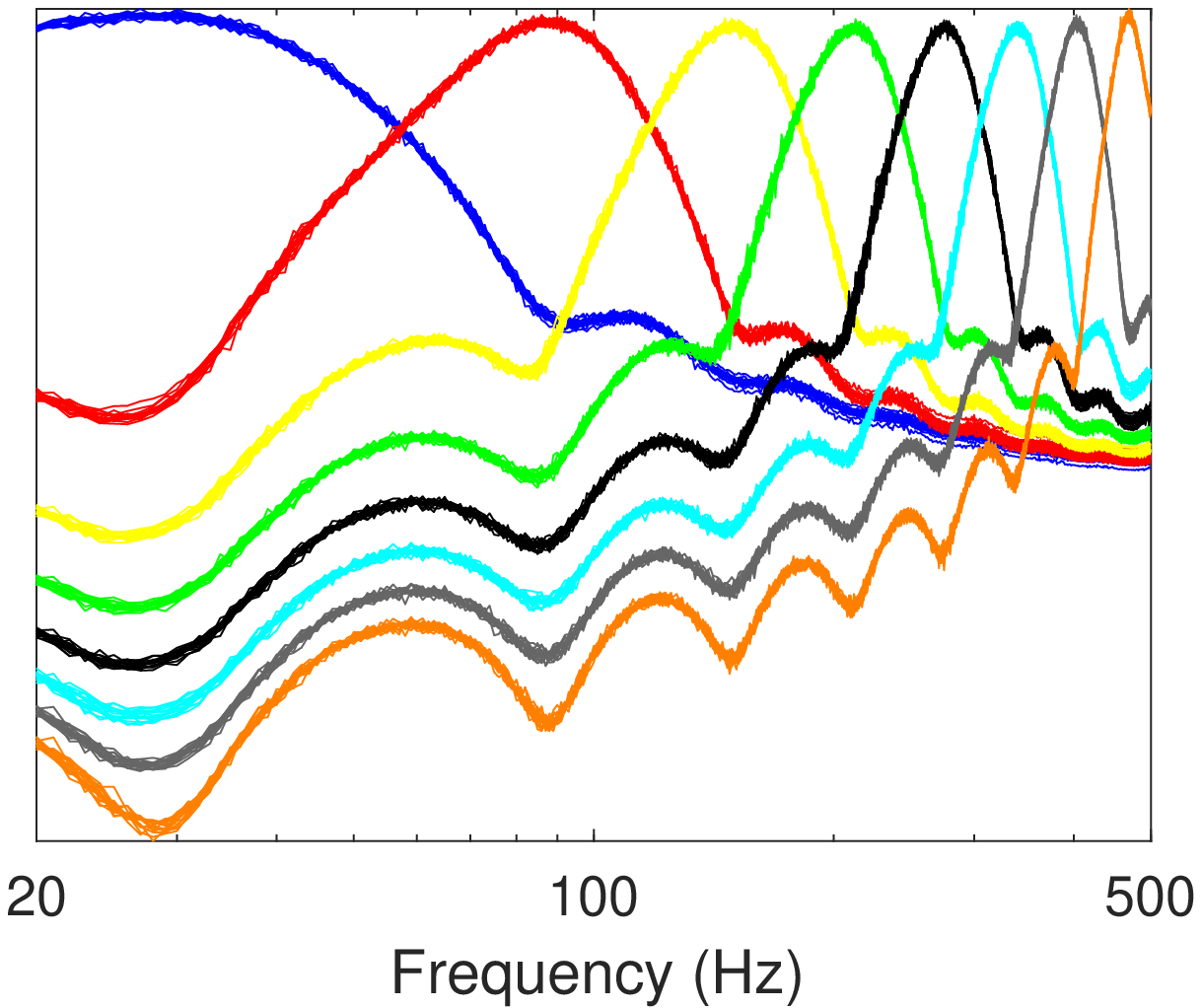}
	\end{subfigure}
	
	\begin{subfigure}{0.25\textwidth}
		\centering
		\includegraphics[trim={5mm 0mm 8mm 5mm},clip,width=1.0\linewidth]{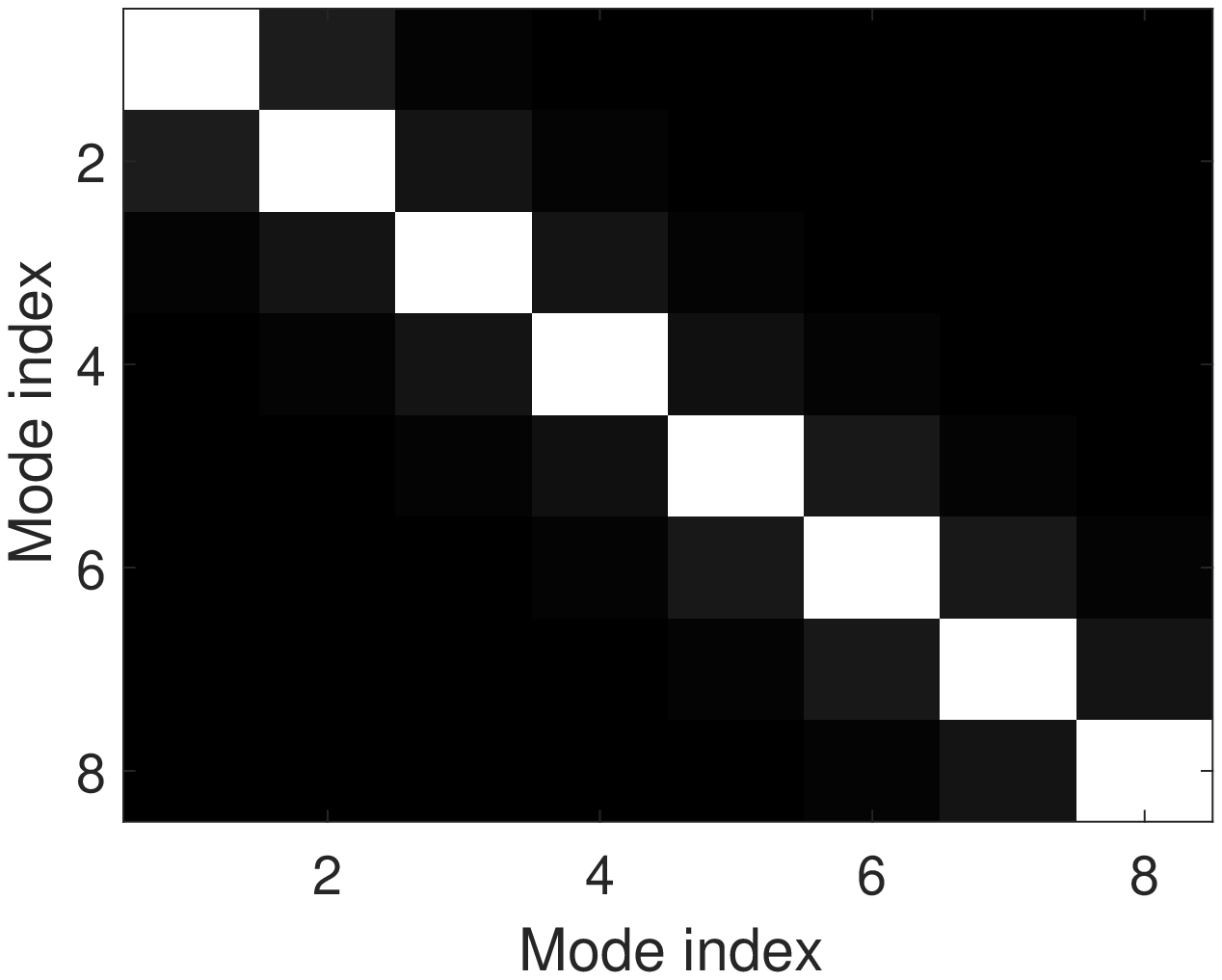}
	\end{subfigure}
	\hspace{-3mm}
	\begin{subfigure}{0.25\textwidth}
		\centering
		\includegraphics[trim={5mm 0mm 8mm 5mm},clip,width=1.0\linewidth]{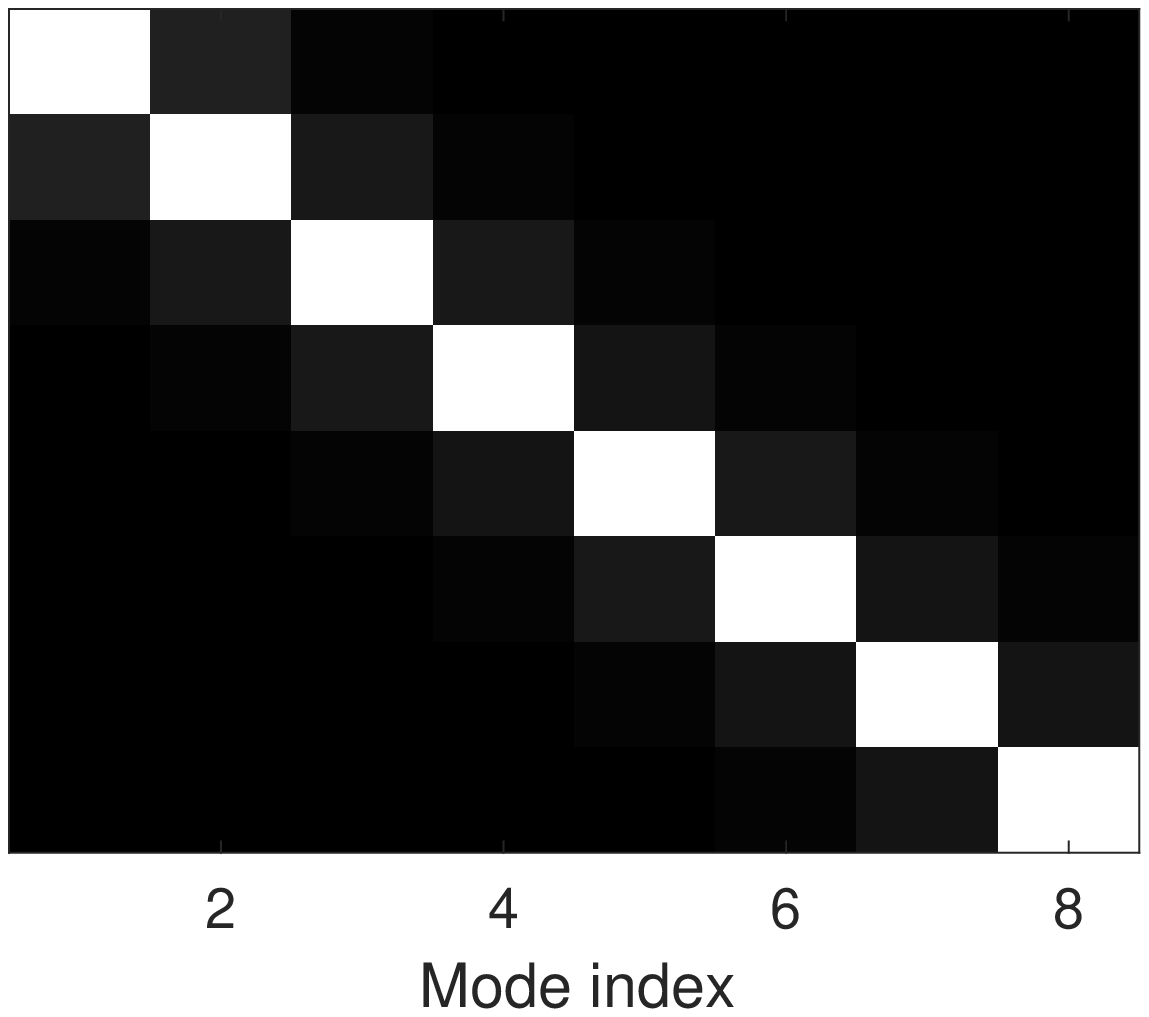}
	\end{subfigure}
	\hspace{-3mm}
	\begin{subfigure}{0.25\textwidth}
		\centering
		\includegraphics[trim={5mm 0mm 8mm 5mm},clip,width=1.0\linewidth]{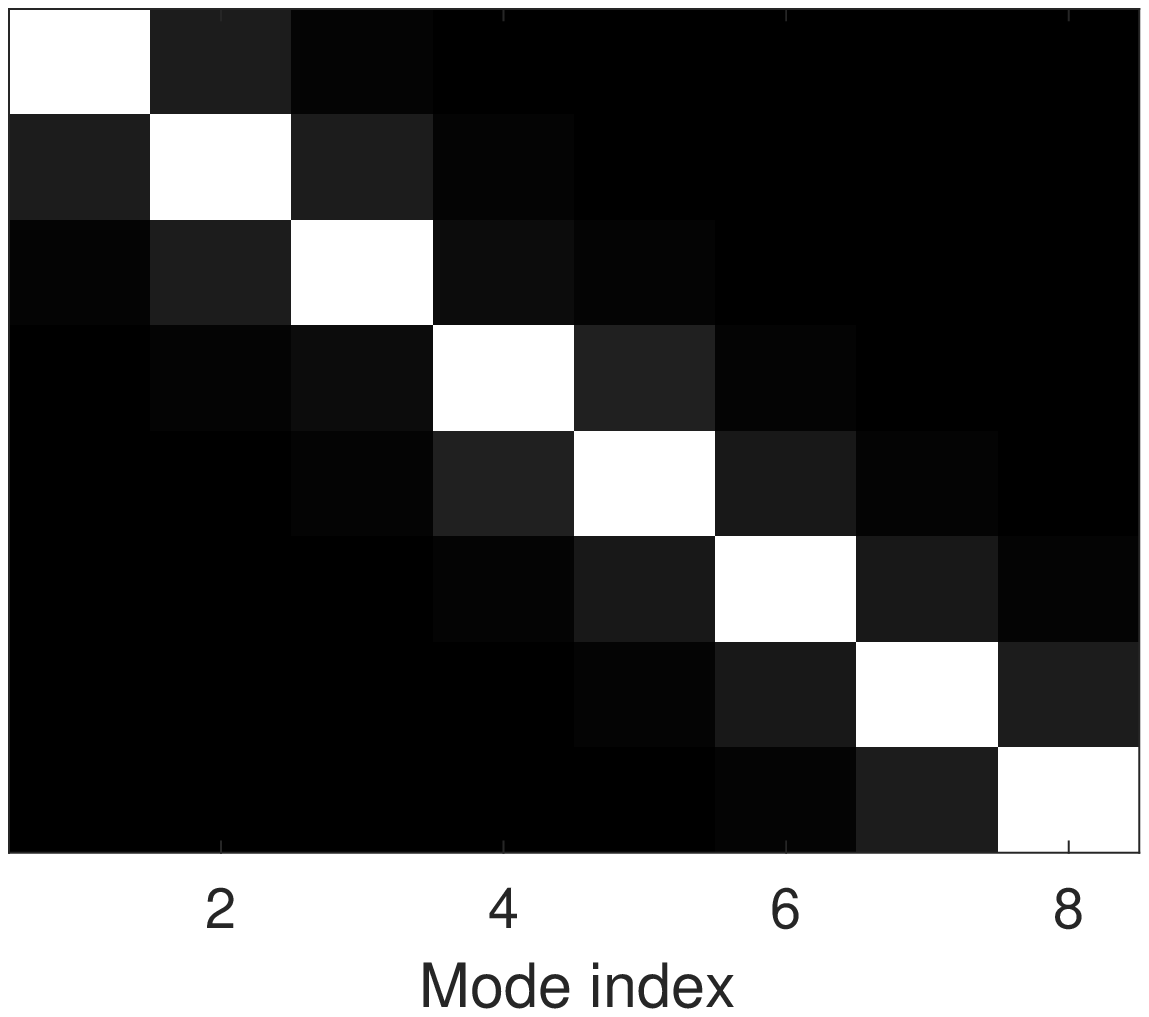}
	\end{subfigure}
	\hspace{-3mm}
	\begin{subfigure}{0.25\textwidth}
		\centering
		\includegraphics[trim={5mm 0mm 8mm 5mm},clip,width=1.0\linewidth]{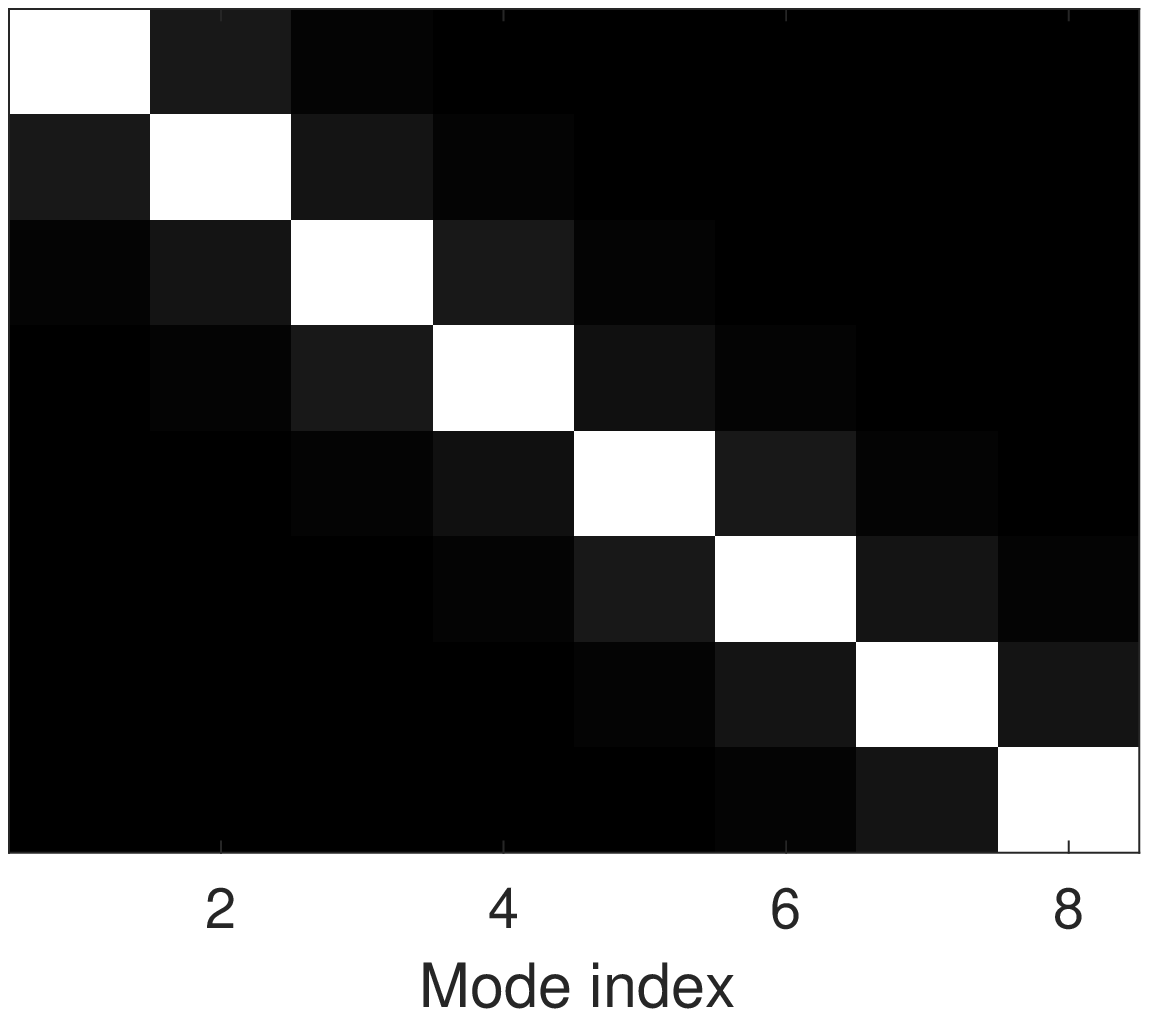}
	\end{subfigure}
	
	\caption{MVMD decomposition of wGn for increasing data channels. (Top row) Illustration of the Filterbank structure and mode-alignment of MVMD decomposition for increasing data channels i.e., $C=2,\mbox{ }4,\mbox{ }8,\mbox{ }16$ from left to right respectively. (Second row) Illustration of quasi-orthogonality of MVMD modes for increasing data channels i.e., $C=2,\mbox{ }4,\mbox{ }8,\mbox{ }16$ from left to right respectively. Increasing number of data channel did not affect the performance of the algorithm in terms of filterbank structure, mode-alignment and   quasi-orthogonality of modes.}
	\label{fig:varyC}
\end{figure*}

\subsection{Noise robustness}

One of the highlights of the VMD related algorithms is its robustness to noise. The origin of this robustness is the inherent flexibility of the method to forgo strict reconstruction property in the presence of noise. 
We demonstrate here that MVMD also inherits that property from standard VMD as a result of being a natural extension of the method. In the case of MVMD, we achieve that by making the Lagrangian term in \eqref{eq:mvmd_lang} to vanish by forcing the parameter $\tau$ in \eqref{eq:mvmd_lambda} to be equal to zero. In this section, our interest specifically is in showing that the mode-alignment property of MVMD is preserved by the presence of noise in input channels. We also compare the performance of our method with MEMD which is known to be very sensitive to noise in input data channels. 

For our experiments, we obtained a bivariate test signal that consisted of a mixture of three tones in its two channels. The 2-Hz and 288-Hz tone were present in both the channels whereas 24-Hz tone was only present in the channel-2. We added an unbalanced wGn with zero mean and variances of $\sigma^2=0.1$ and $\sigma^2=0.05$ that resulted in the SNR of $-dB$ and $-dB$ in channel-1 and channel-2 respectively. The resulting decomposed modes obtained from MVMD are shown in Figure \ref{fig:noisemode}(a). Notice that the MVMD recovers the common modes of 2-Hz and 288-Hz tones in both channels in mode 1 and mode 3 respectively. The 24-Hz tone was only present in channel-2 of mode 2. We also show the decomposed intrinsic mode functions (IMFs) obtained by applying the MEMD algorithm to the same data set. A visual comparison between the two sets of modes reveal that MVMD is more robust to noise in that the mode-alignment is more prominent across MVMD modes as compared to those obtained from MEMD. Some leakage of information can be seen in IMF 4 across channels and there is definite mode-mixing within the first three IMFs: $c_3$ does not fully represent the 288-Hz tone since the first two IMFs also share that information i.e., mode-mixing which may be problematic in practical applications. Similar results were reported in \cite{bib:vmd} while comparing single channel VMD vs EMD in terms of robustness to noise. 

By design, the VMD and MVMD algorithms are inherently robust to noise  due to control over Lagrangian multipliers via the parameter $\tau$. In the case of noisy input data, it may not be desirable to obtain a decomposition that fully reconstructs the noisy input data. That can be controlled within VMD and MVMD by forcing the parameter $\tau$ to be equal to zero. In MEMD, on the other hand, there is no mechanism to stop the noise entering into the decomposition process. As a result, in many practical application involving noisy data, a common preprocessing step in EMD involves rejecting noisy IMFs manually which may be prone to errors.

\subsection{Performance evaluation for increasing number of channels}
The performance of MVMD is evaluated for varying number of data channels. In our experiments, the input were 200 realizations of wGn process with $C=2$, $4$ ,$8$ and $16$ noise channels. After obtaining the MVMD modes in each case, we investigated how the mode-alignment property, filterbank structure and quasi-orthogonality of the modes were affected with increasing number of input channels. We show the resulting filterbank plots and the correlation coefficient matrices in Figure \ref{fig:varyC}. It can be observed from the top row of the Figure \ref{fig:varyC} that the filterbank and mode-alignment properties within MVMD are perfectly retained as the number of channels are increased. Similarly, quasi-orthogonality of MVMD decomposition is also not affected by increasing number of channels as highlighted by the diagonal nature of all correlation plots shown in the second row of Figure \ref{fig:varyC}.            

\subsection{Real World Examples}

The utility of the proposed MVMD method in signal decomposition and T-F analysis of real world data is illustrated via two examples. Those include separation of $\alpha$-rhythms in multivariate EEG data and decomposition of fetal heart rate (FHR) and maternal uterine contraction (UC) recordings from cardiotocographic (CTG) traces \cite{bib:saleem}. 
Both these examples involve multiscale signal decomposition of multivariate data and require alignment of similar frequency content across multiple channels of each decomposed mode. MVMD, by its design, achieves such a decomposition and is therefore a suitable candidate for such applications. 

\begin{figure}[!t]
\centering
\begin{subfigure}{0.5\textwidth}
		\includegraphics[trim={2mm 9mm 0mm 0mm},clip,width=1.0\linewidth,height=0.34\linewidth]{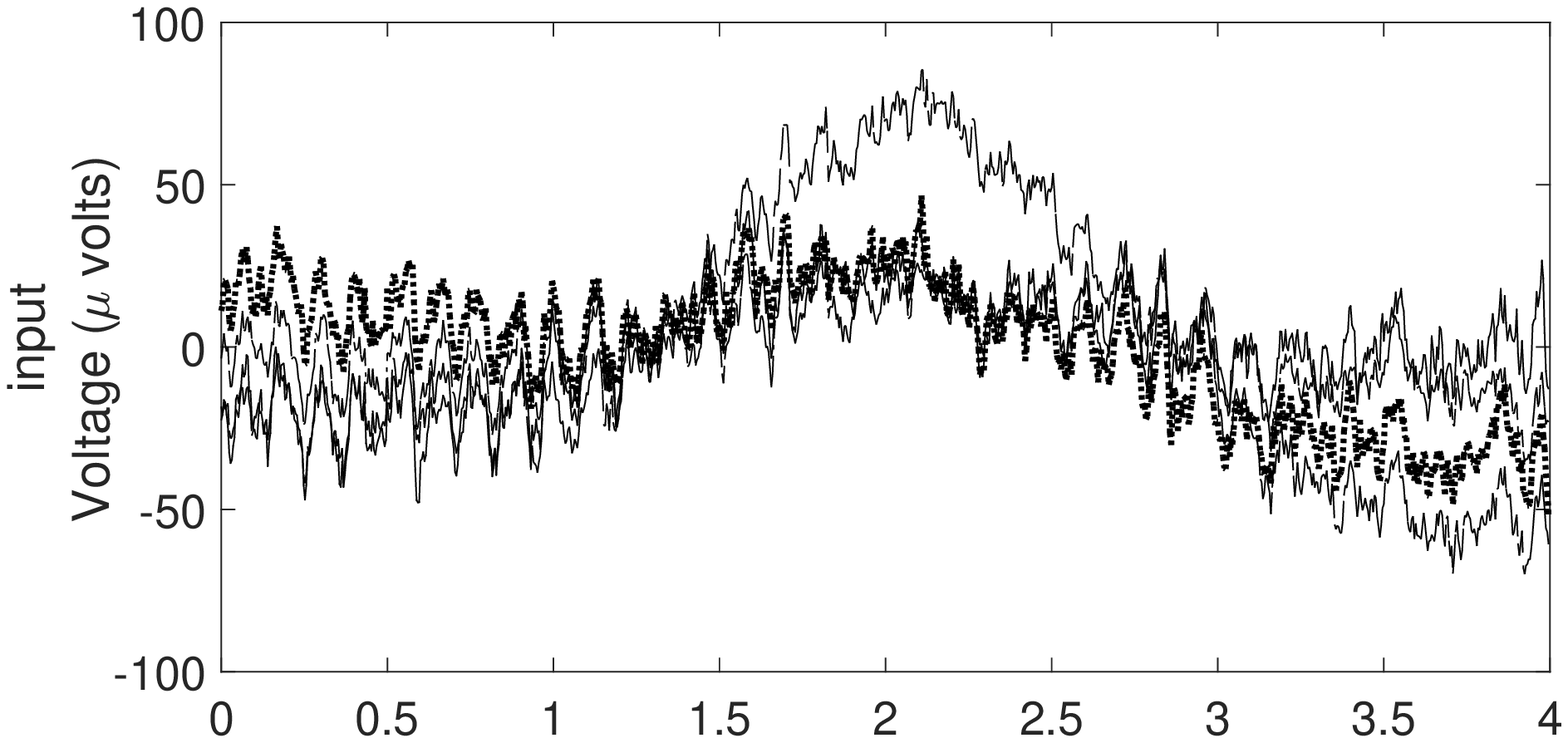}
		\includegraphics[trim={2mm 8mm 0mm 0mm},clip,width=1.0\linewidth,height=0.34\linewidth]{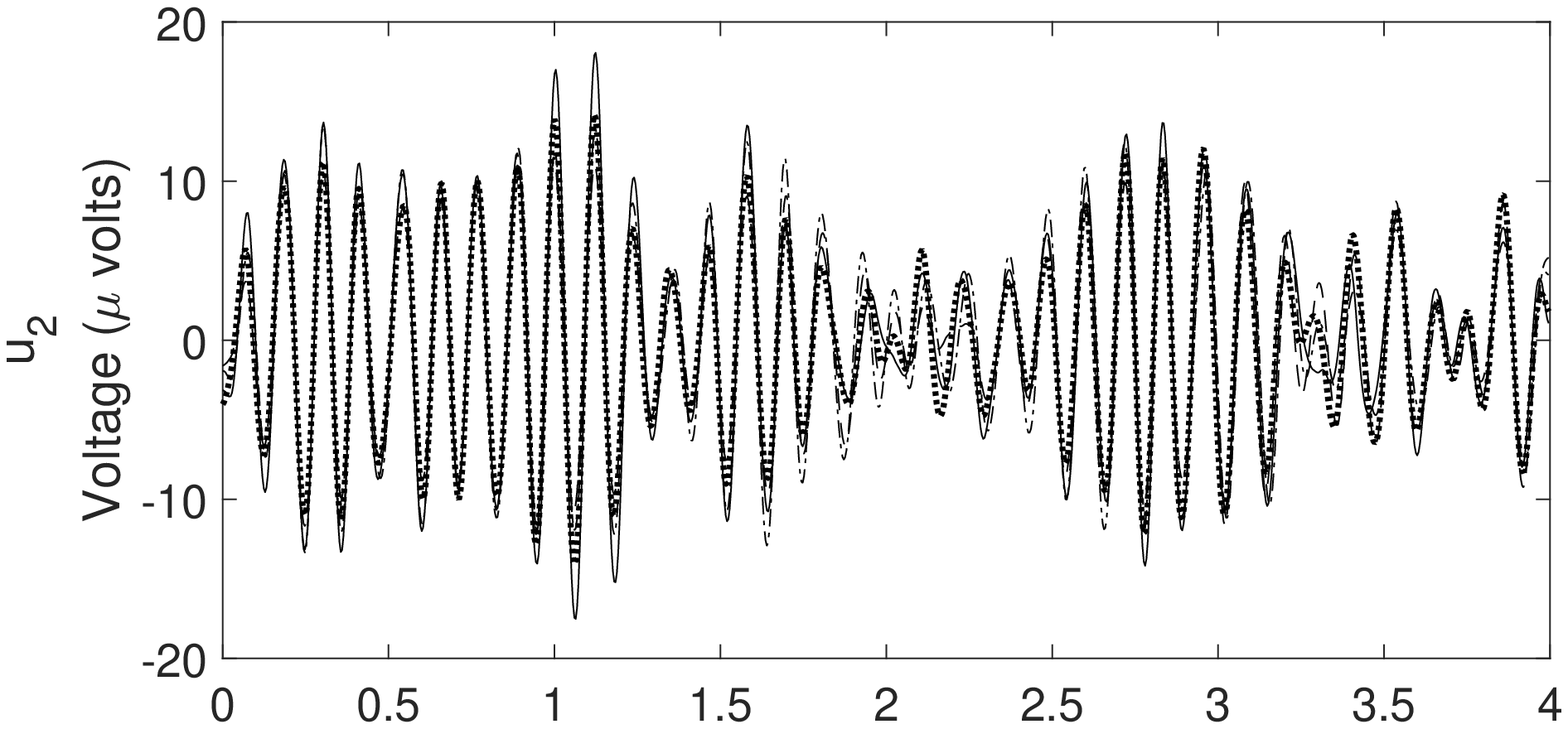}
		\includegraphics[trim={2mm 8mm 0mm 0mm},clip,width=1.0\linewidth,height=0.34\linewidth]{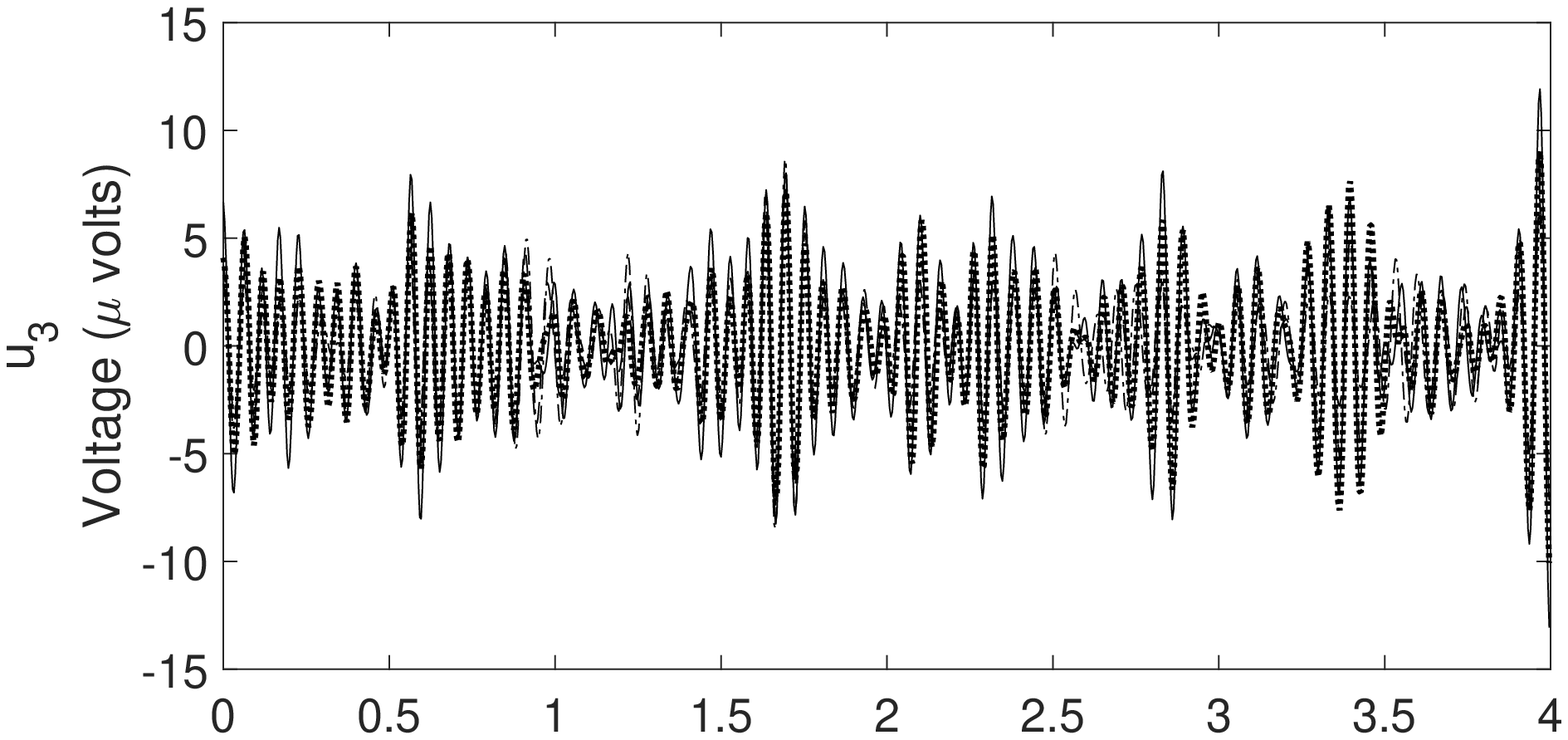}
		\includegraphics[trim={1mm 0mm 0mm 0mm},clip,width=1\linewidth,height=0.39\linewidth]{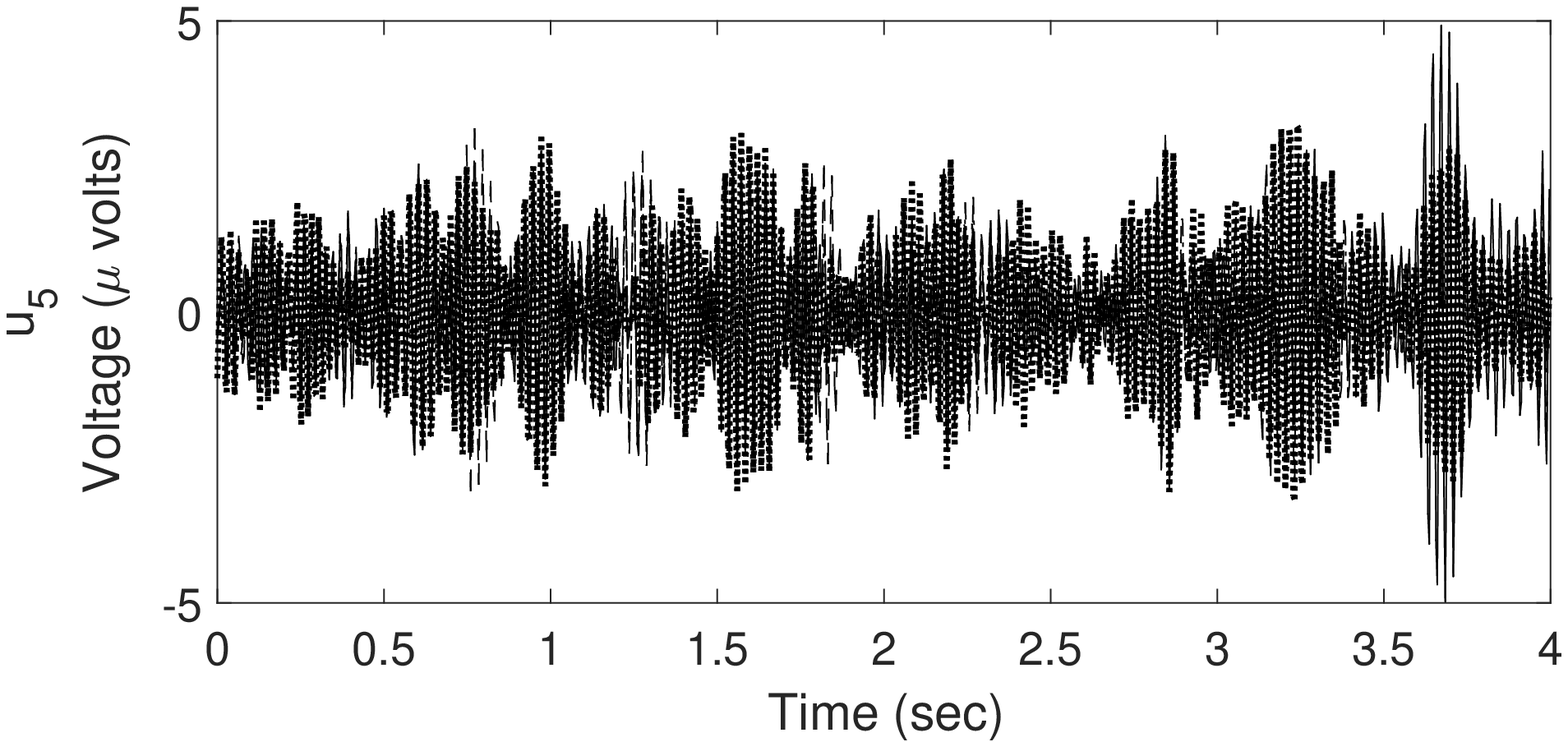}
\caption{}	
\end{subfigure}
	
\begin{subfigure}{0.5\textwidth}
		\includegraphics[trim={2mm 0mm 9mm 5mm},clip,width=0.5\linewidth]{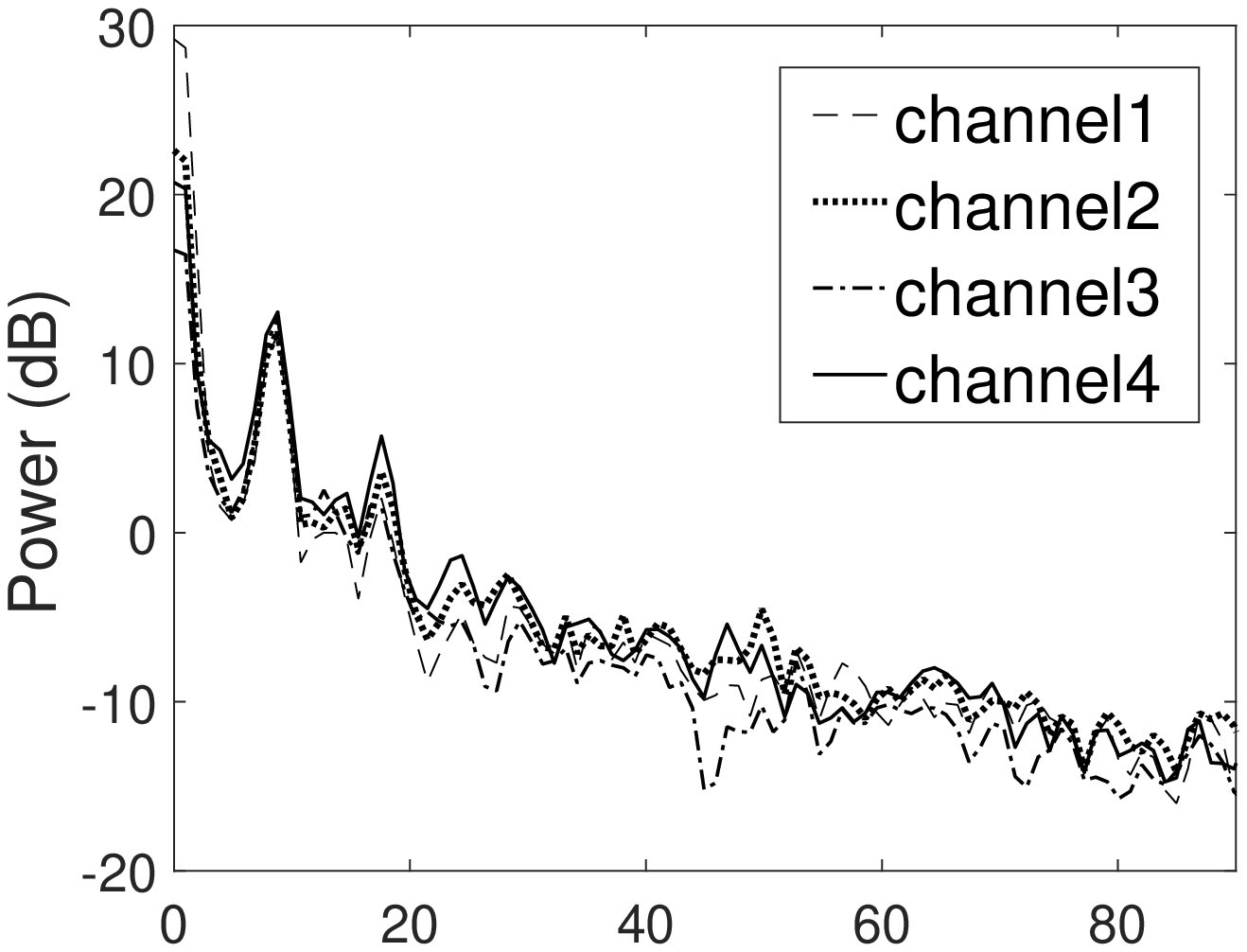}\hspace{-4mm}
		\includegraphics[trim={2mm 0mm 9mm 5mm},clip,width=0.5\linewidth]{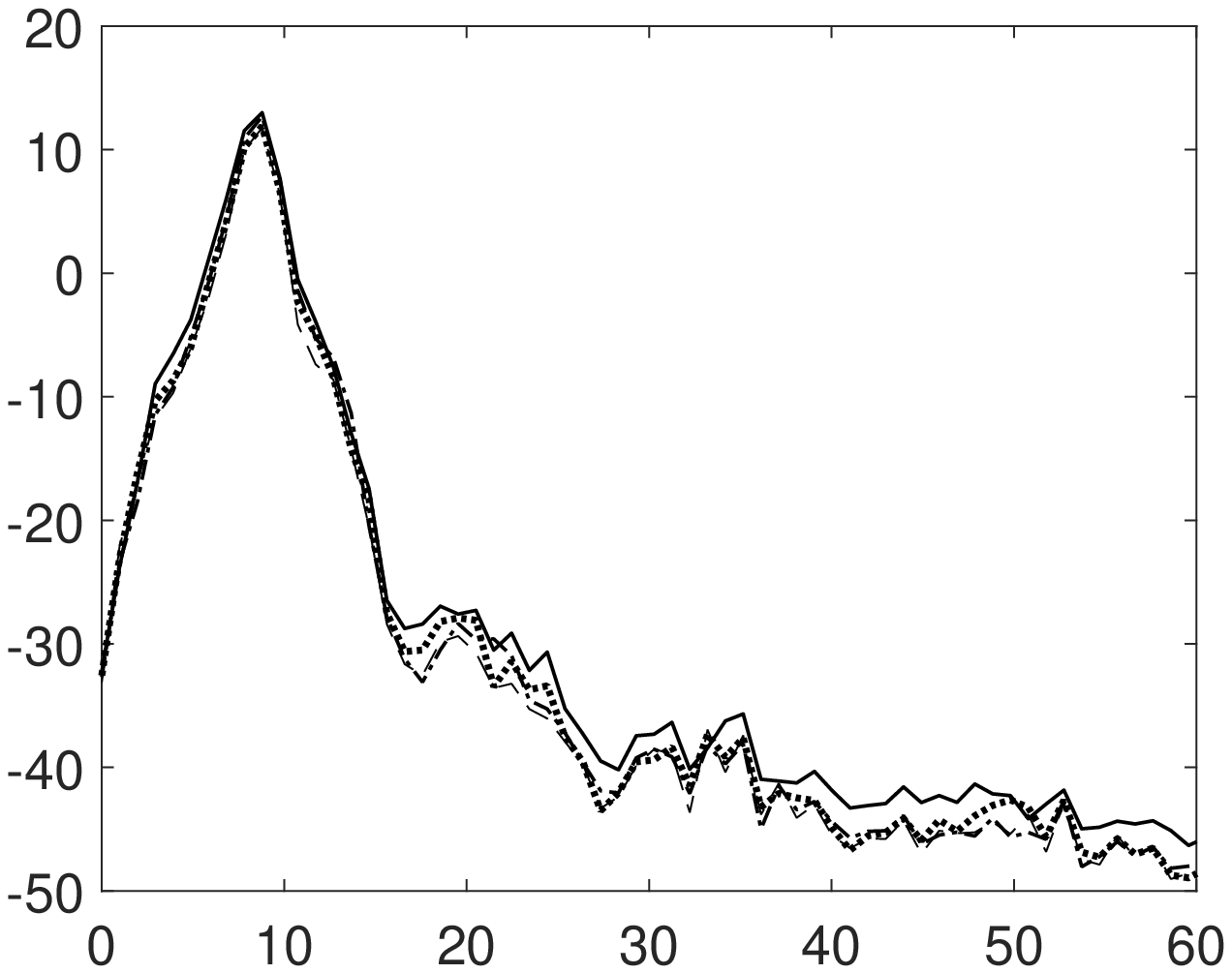}\hspace{-2mm}
		\includegraphics[trim={1mm 0mm 9mm 2mm},clip,width=0.5\linewidth]{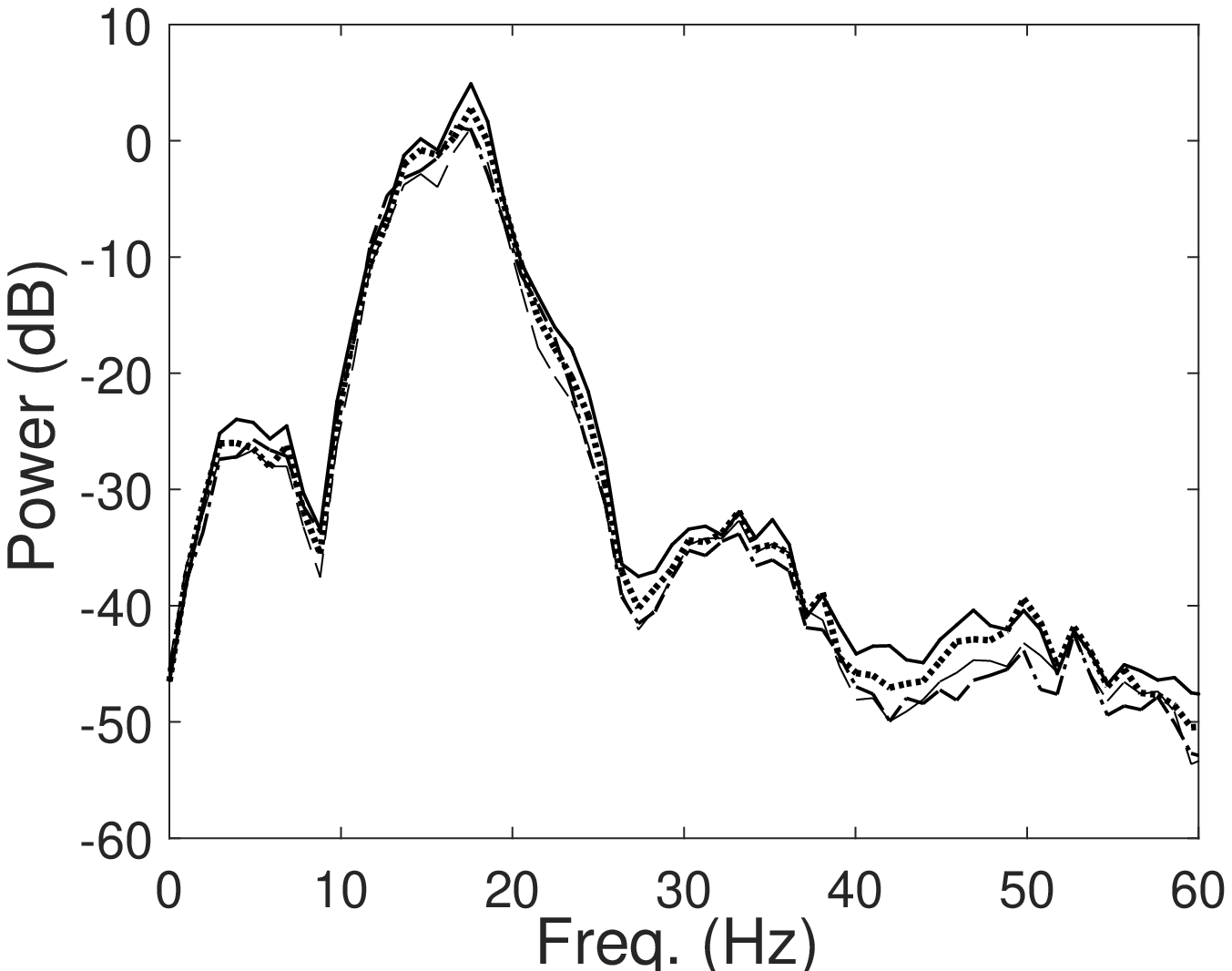}\hspace{-2mm}
		\includegraphics[trim={2mm 0mm 9mm 2mm},clip,width=0.5\linewidth]{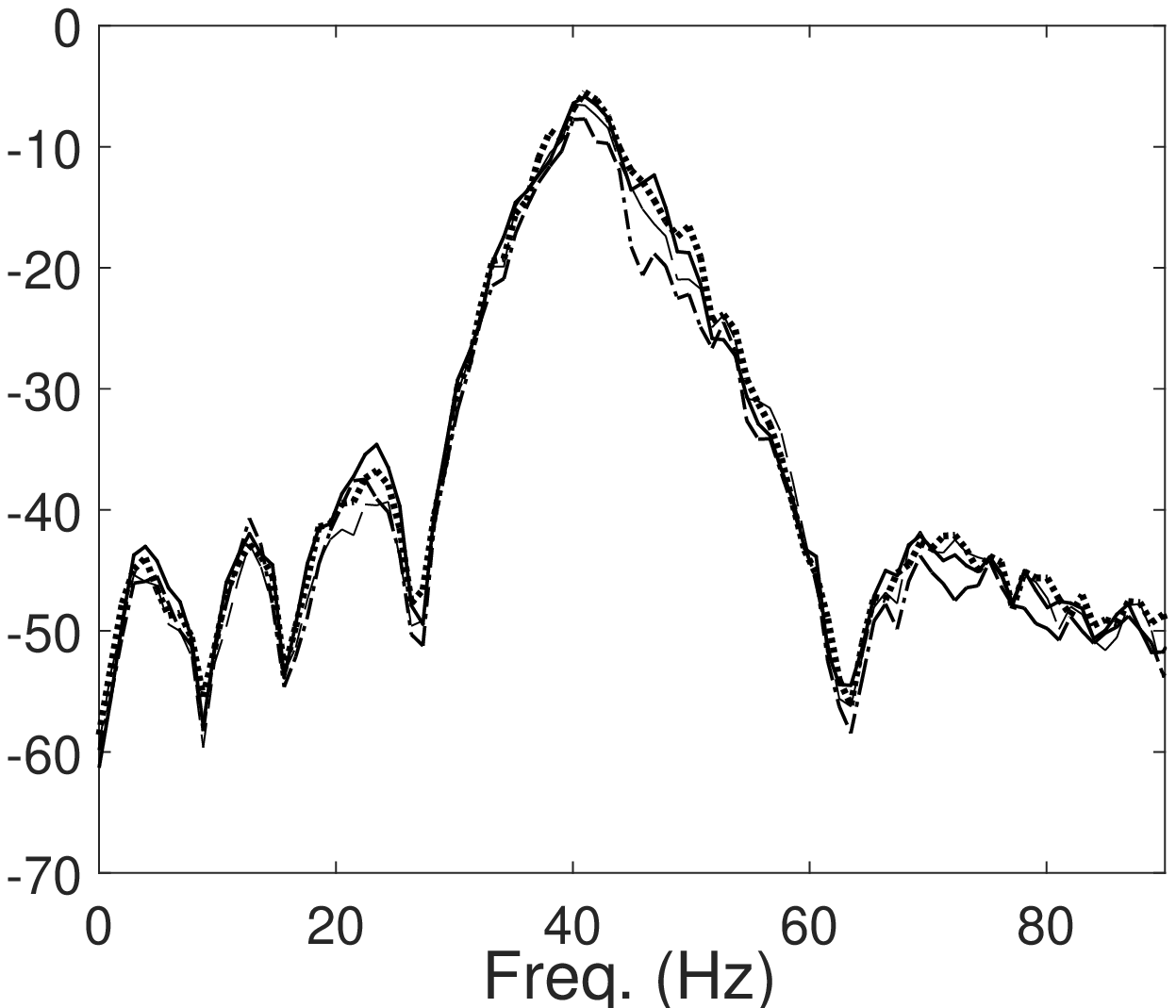}
	\caption{}

\end{subfigure}
		
	\caption{MVMD decomposition of a 4-channel EEG data; (a) Time plots of input data along with selected modes from MVMD ($u_2$, $u_3$, and $u_5$); (b) Smoothed PSD estimates of input (top left) and selected modes ($u_2$ (top right), $u_3$ (lower left), and $u_5$ (lower right).}
	\label{fig:MDD}
\end{figure}

\subsubsection{Separation of $\alpha$ rhythms in EEG}

An example of multichannel Electroencephalogram (EEG) signal processing for brain computer interface (BCI) using MVMD is described. A 4-channel EEG data was obtained in an experiment that involved a subject who remained in the relaxed state with eyes closed for a certain period of time. It is well established that $\alpha$-rhythms, corresponding to frequency range of 8-12 Hz, are detected in EEG data during the relaxed and eyes-closed state. 
The aim here is to investigate the frequency localization property of MVMD to detect $\alpha$-rhythms at multiple channels. The EEG data were obtained through the OpenBCI Cyton board at a sampling frequency of 250 samples per second. 

Figure \ref{fig:MDD} shows the time plots of the original data and selected modes ($u_2$, $u_3$, $u_5$) obtained from MVMD (a) along with their estimated power spectral density (PSD) plots (b). Note that the $\alpha$-rhythm within EEG is localized in the second mode $u_2$; it was also observed that all channels corresponding to $u_2$ contained $\alpha$-rhythm thus highlighting the mode-alignment property of MVMD across multiple channels of the data. Similar mode-alignment was observed across multiple channels of all the extracted modes. $u_5$ corresponds to artifacts due to the AC mains power line. The smoothed PSD plots of the three modes in Figure \ref{fig:MDD}(b), with each showing distinct peak along a certain frequency band, and the alignment of PSD estimates from different channels highlight the mode-mixing and mode-alignment property of MVMD respectively.

\subsubsection{Decomposition of fetal heart rate (FHR) and maternal uterine contraction (UC) recordings from cardiotocographic (CTG) signal}

Fetal heart rate (FHR) signal monitoring provides a useful means to assess the fetal health in obstetric practice. Precise monitoring of FHR can be achieved through Cardiotocography (CTG) that also captures maternal uterine contractions (UCs), making CTG an attractive technique in obstetrics \cite{bib:warrick}. A recent study utilized the mode-alignment property of bivariate extension of EMD (BEMD) to specify robust features and measures for classification of vaginal vs cesarean delivery \cite{bib:saleem}. The data was taken from a freely available CTU-UHB intrapartum cardiotocography database available at Physionet: http://www.physionet.org/physiobank/database/ctu-uhb-ctgdb/ \cite{bib:goldberger}. We highlight here the potential of MVMD to avoid mode-mixing and achieve mode-alignment in a single record of FHR-UC data and compare the results from those obtained from BEMD.

\begin{figure}
	\begin{subfigure}{0.5\textwidth}
		\centering
		\includegraphics[trim={2mm 5.5mm 8mm 4mm},clip,width=1\linewidth]{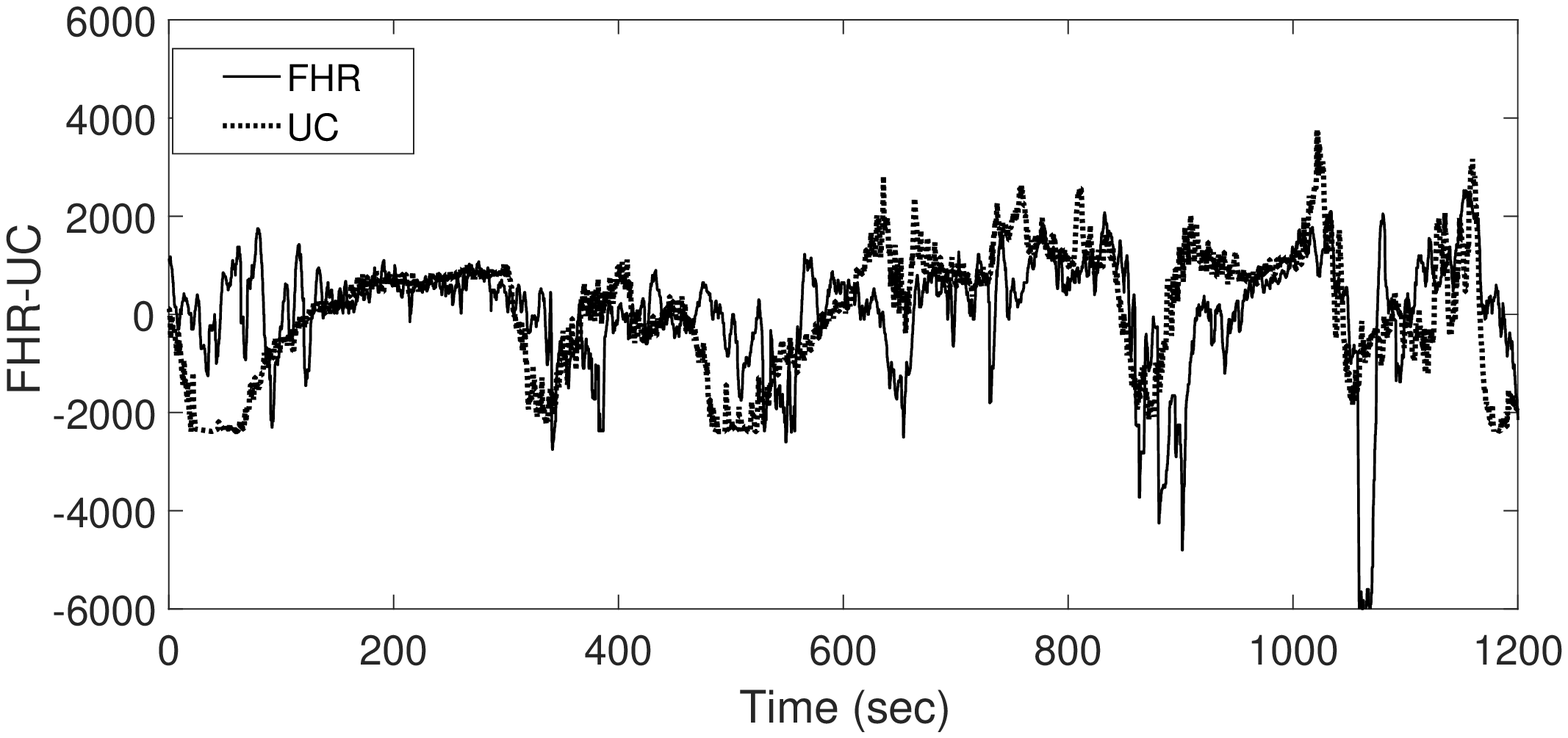}
		\caption{}
	\end{subfigure}
	\begin{subfigure}{0.5\textwidth}
		\centering
		\includegraphics[trim={4mm 0mm 2mm 0mm},clip,width=0.51\linewidth]{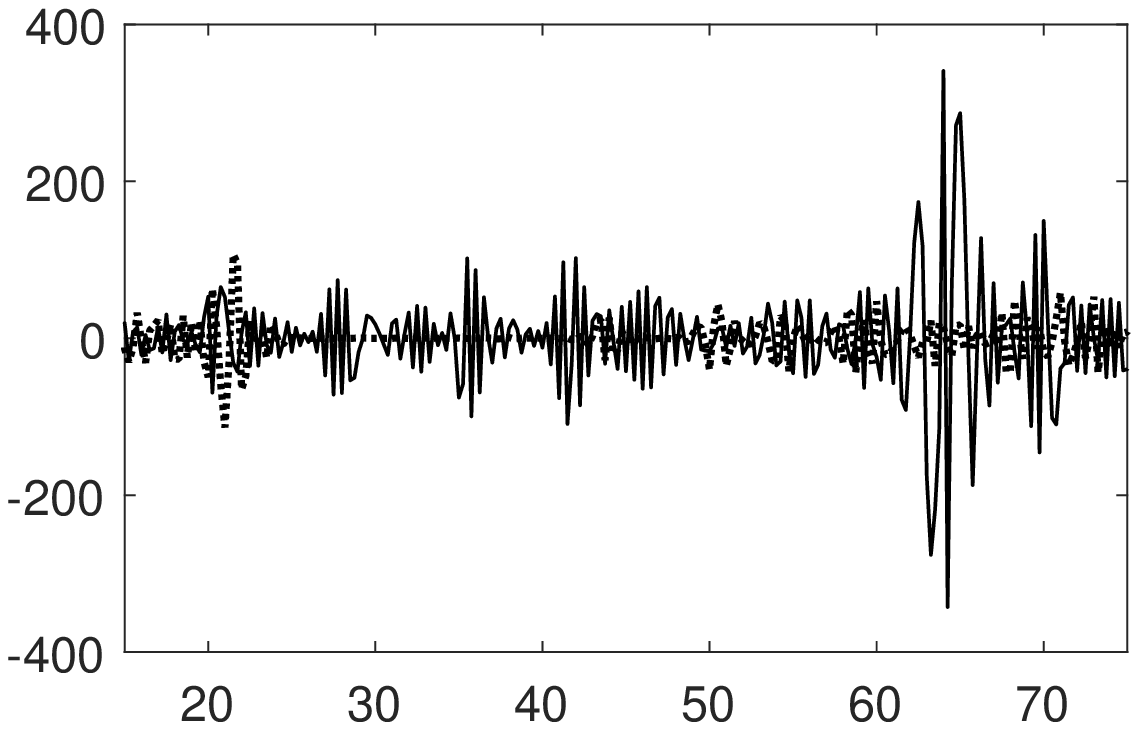}\hspace{-3mm}
		\includegraphics[trim={4mm 0mm 5mm 0mm},clip,width=0.5\linewidth]{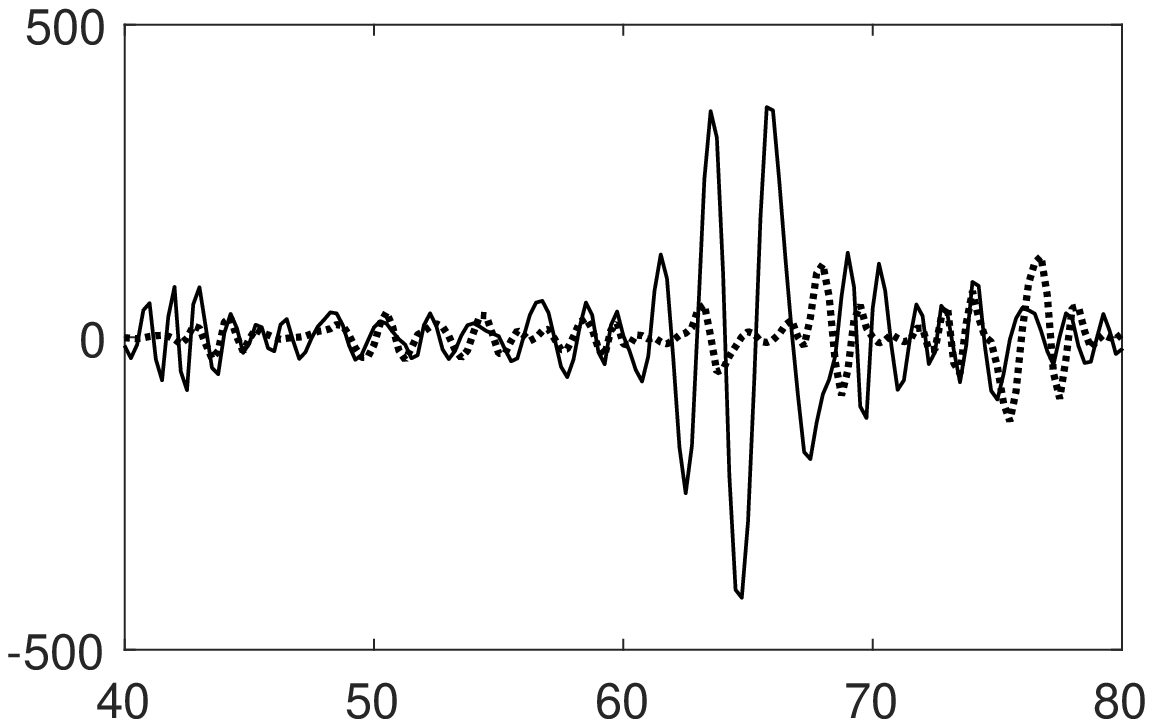}\\
		\includegraphics[trim={4mm 0mm 2mm 0mm},clip,width=0.51\linewidth]{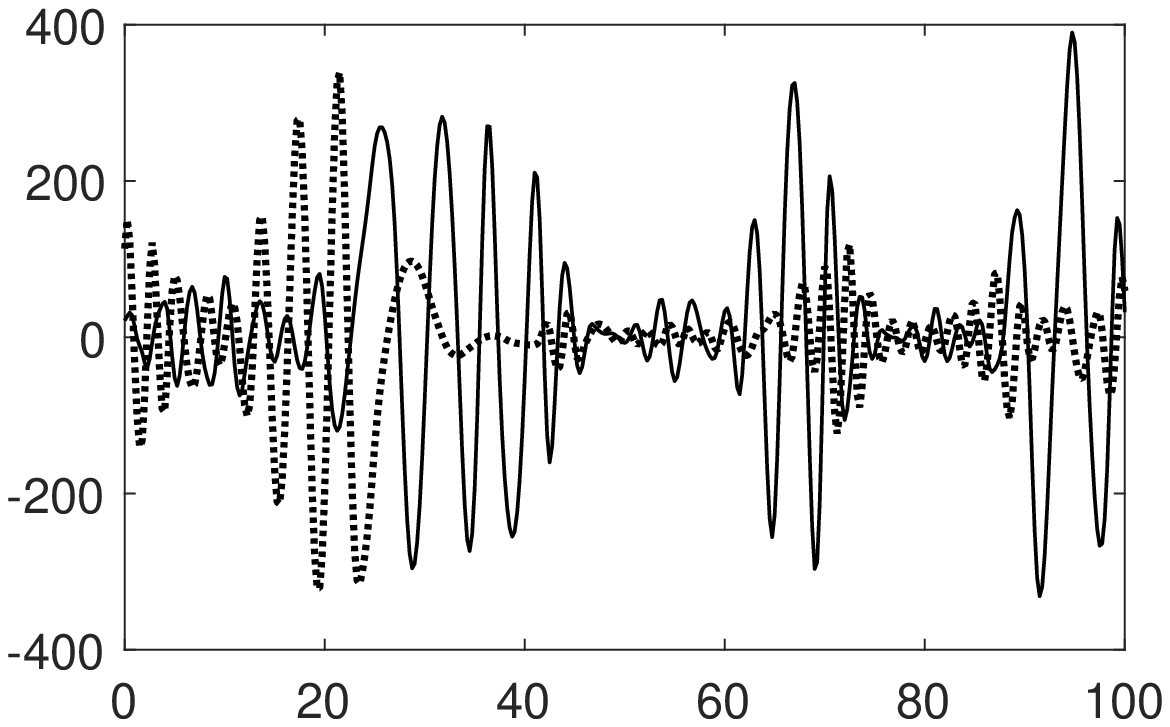}\hspace{-3mm}
		\includegraphics[trim={4mm 0mm 5mm 0mm},clip,width=0.5\linewidth]{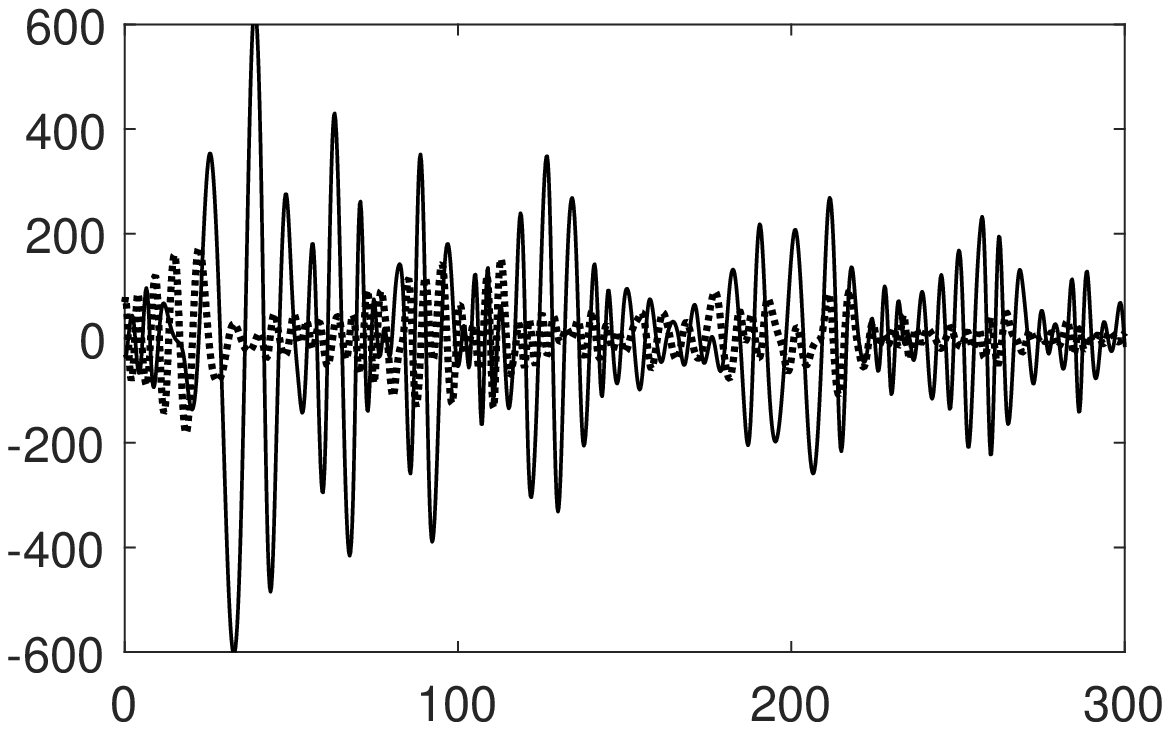}\\
		\includegraphics[trim={4mm 0mm 2mm 0mm},clip,width=0.51\linewidth]{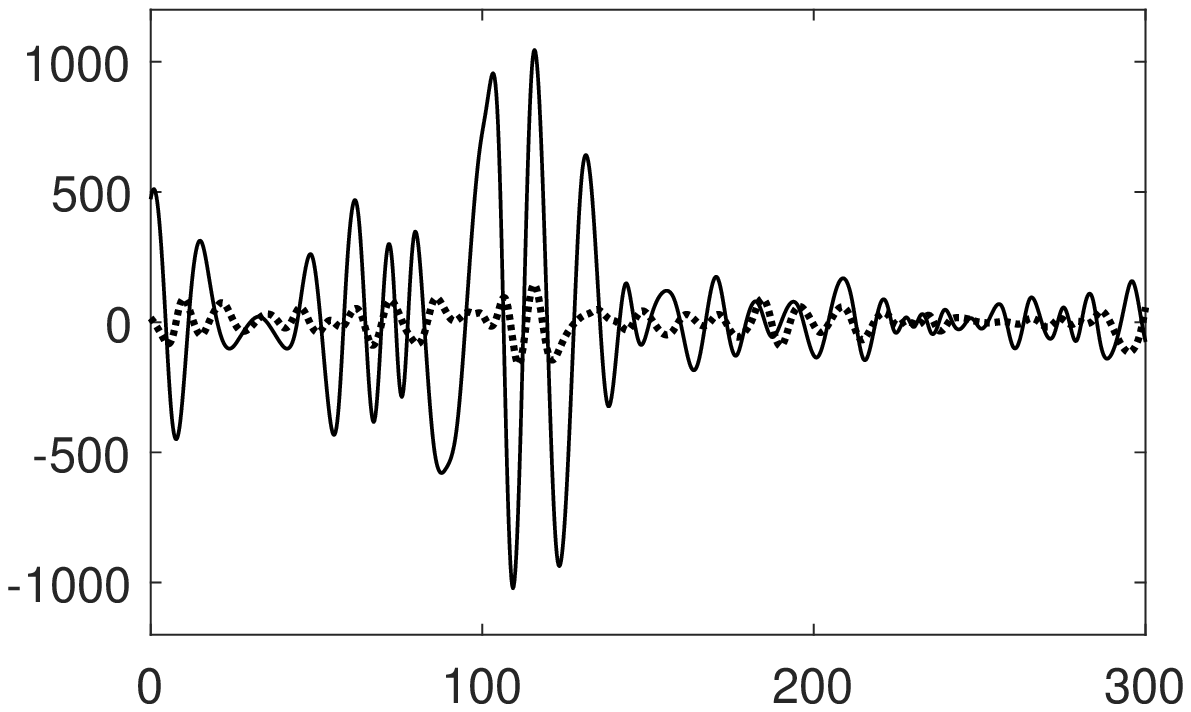}\hspace{-3mm}
		\includegraphics[trim={4mm 0mm 5mm 0mm},clip,width=0.5\linewidth]{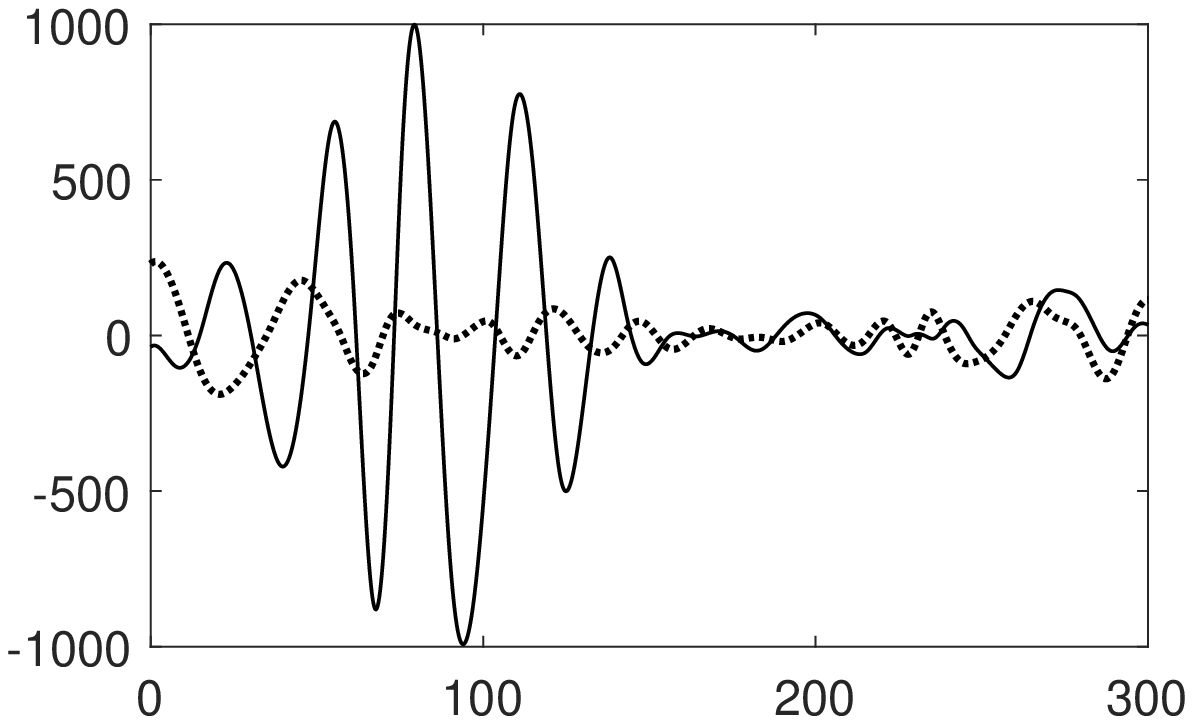}\\
		\caption{BEMD decomposition (Left to right, top to bottom: $c_1$-$c_6$)}
	\end{subfigure}
	\begin{subfigure}{0.5\textwidth}
		\centering
		\includegraphics[trim={4mm 0mm 2mm 0mm},clip,width=0.51\linewidth]{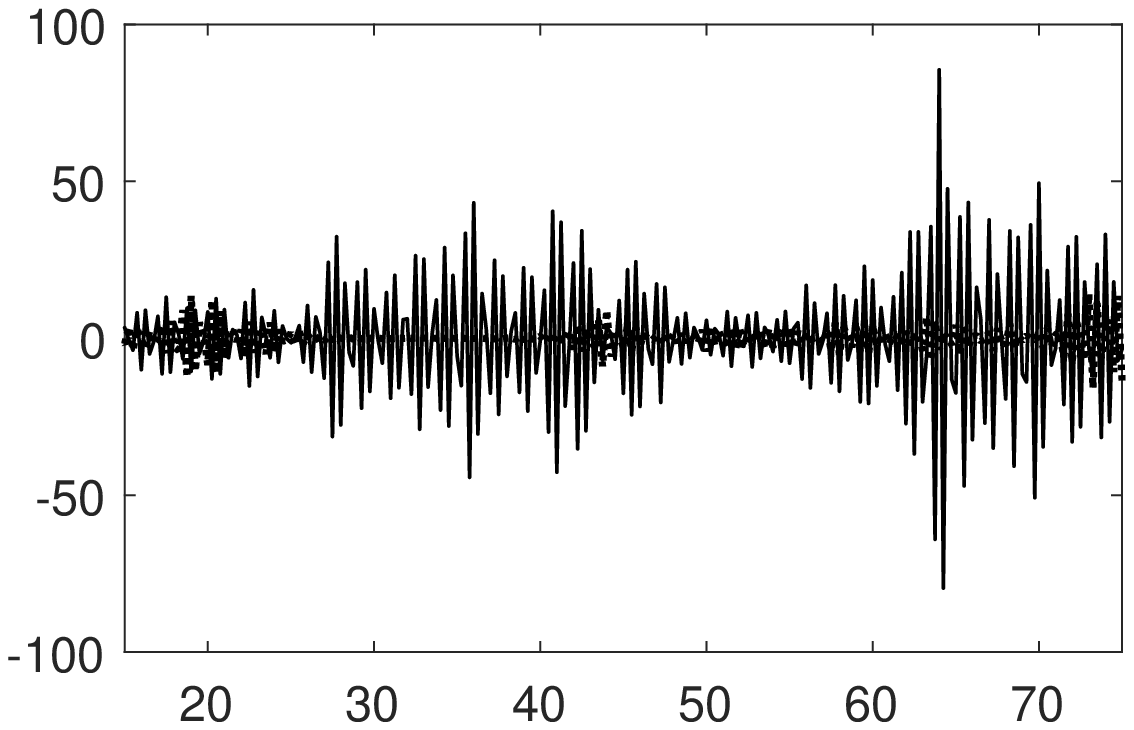}\hspace{-3mm}
		\includegraphics[trim={4mm 0mm 5mm 0mm},clip,width=0.5\linewidth]{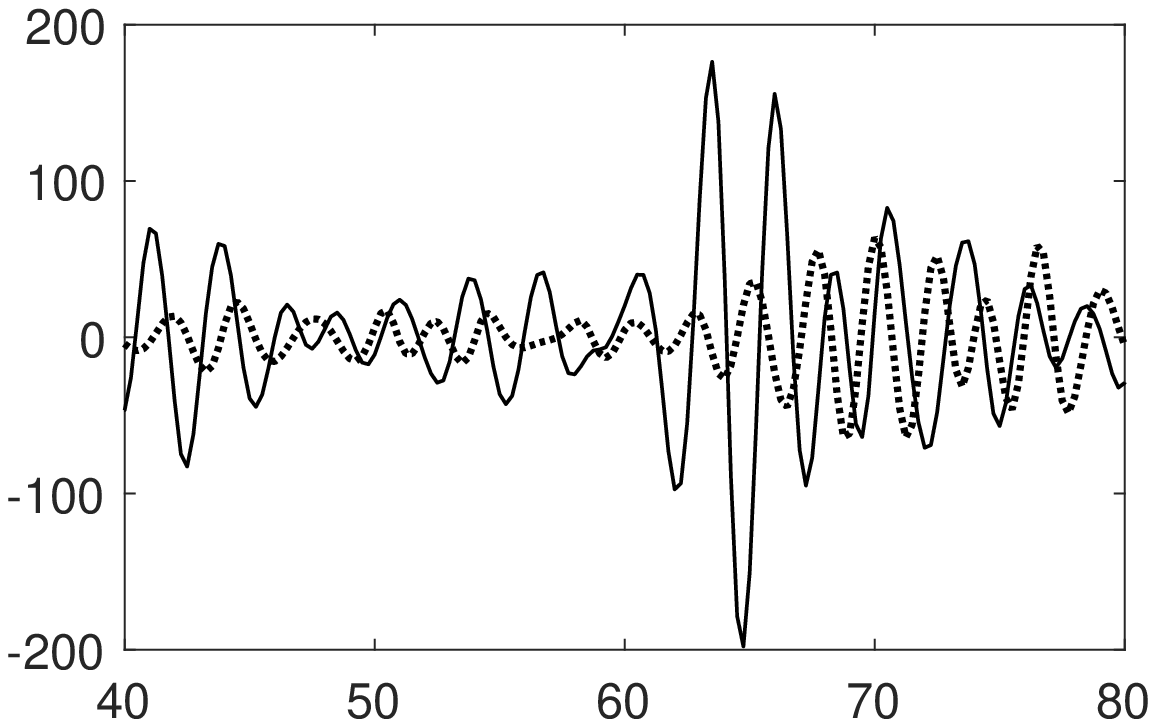}\\
		\includegraphics[trim={4mm 0mm 2mm 0mm},clip,width=0.51\linewidth]{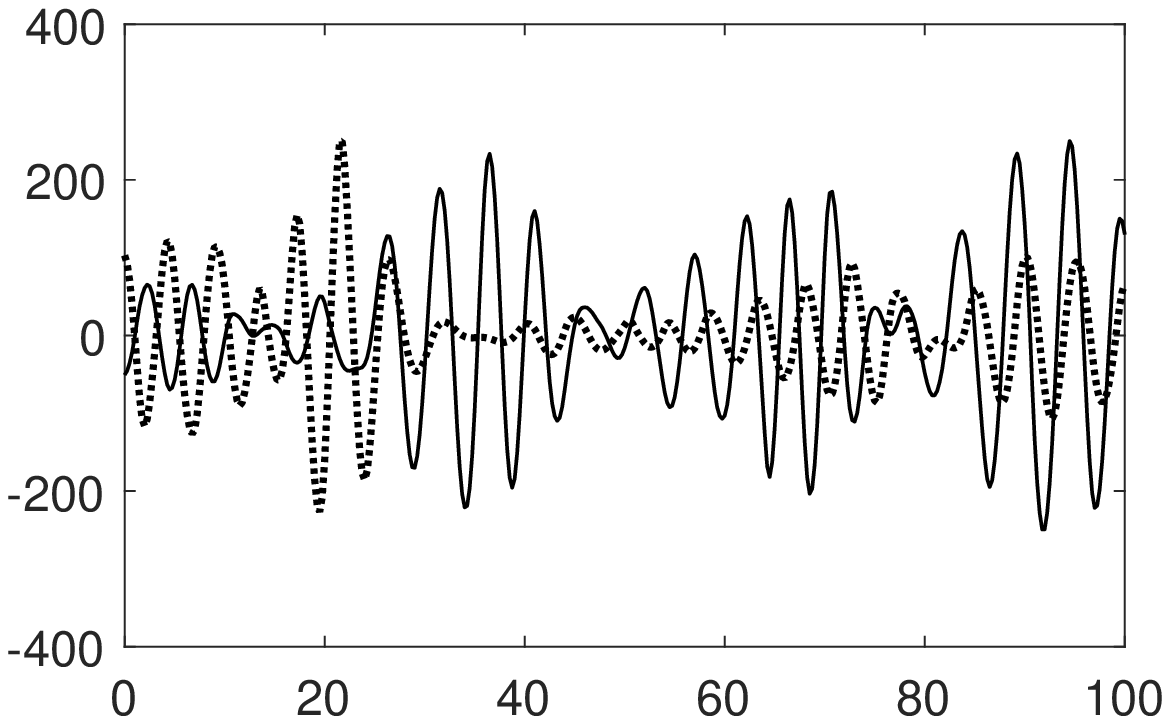}\hspace{-3mm}
		\includegraphics[trim={4mm 0mm 5mm 0mm},clip,width=0.5\linewidth]{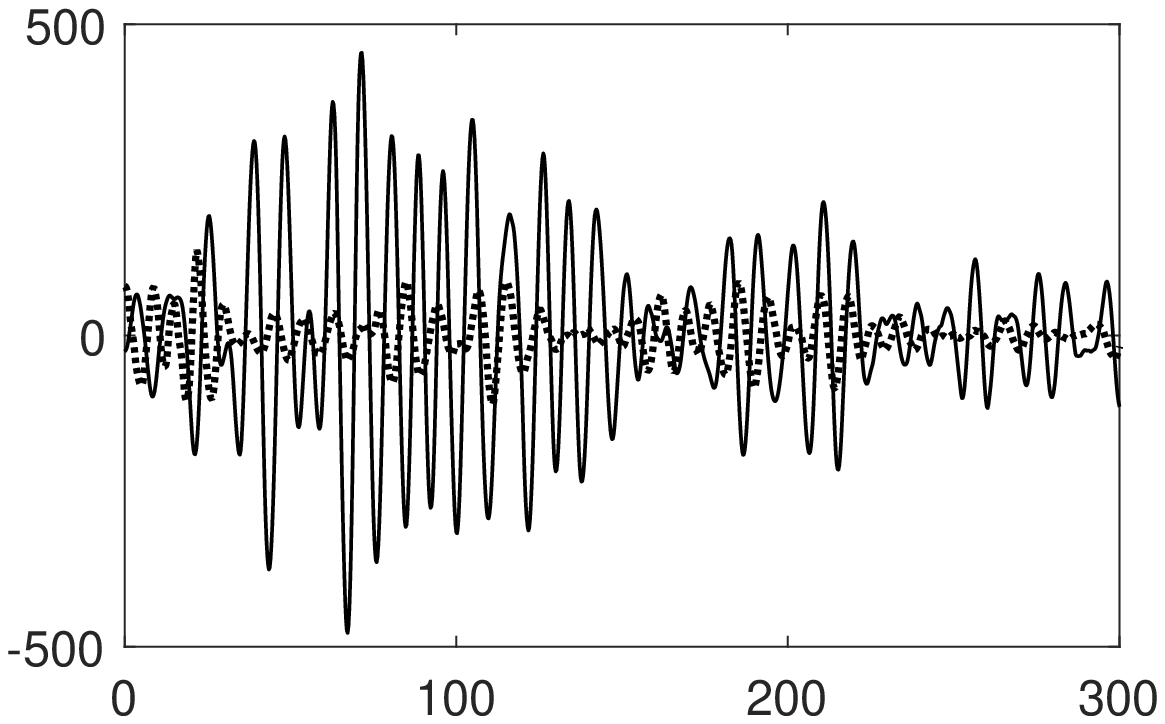}\\
		\includegraphics[trim={4mm 0mm 2mm 0mm},clip,width=0.51\linewidth]{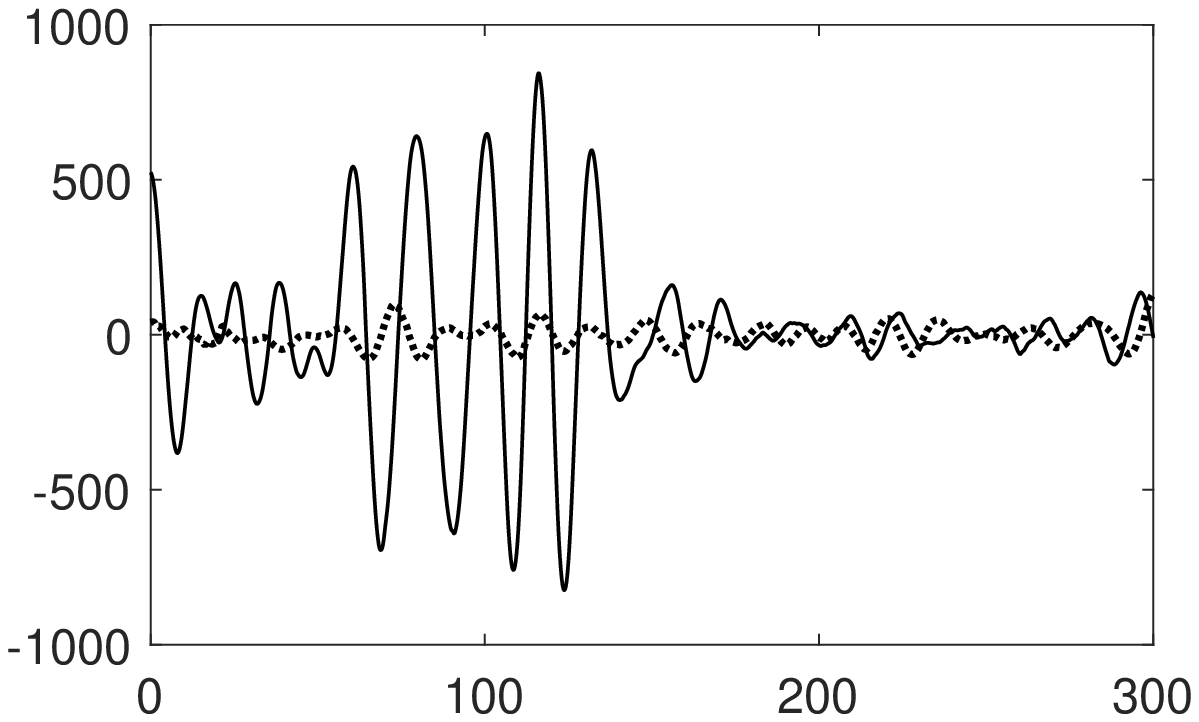}\hspace{-3mm}
		\includegraphics[trim={4mm 0mm 5mm 0mm},clip,width=0.5\linewidth]{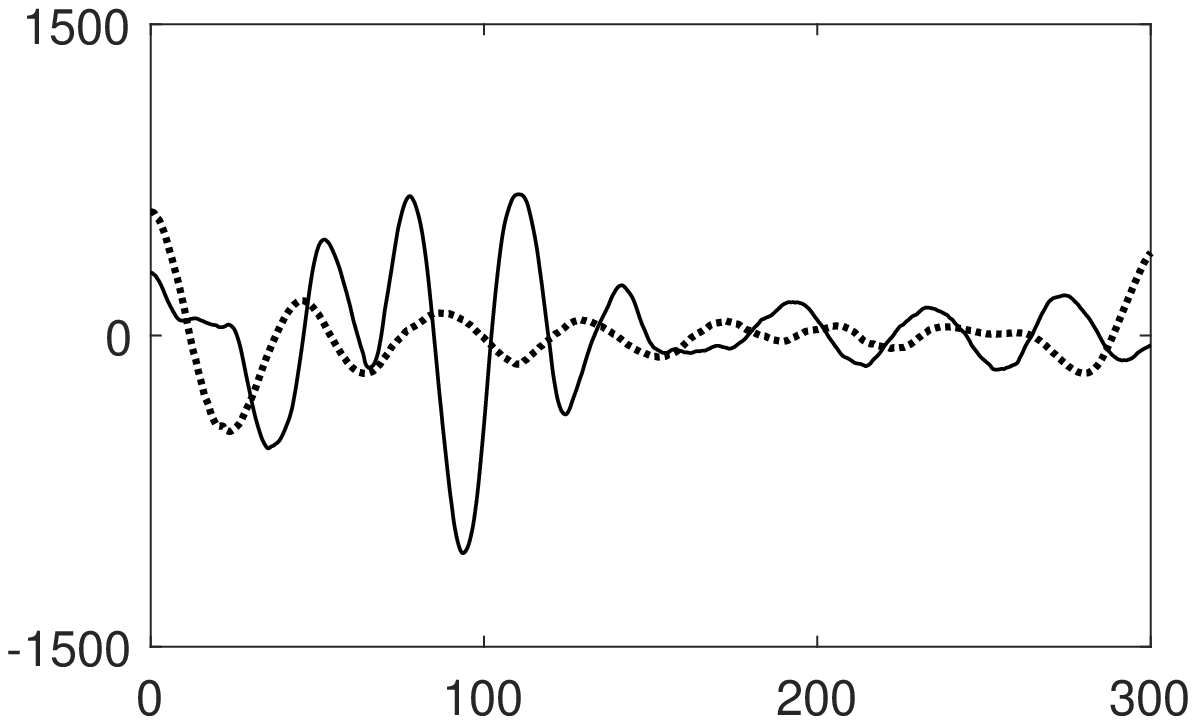}\\
		\caption{MVMD decomposition (Left to right, top to bottom: $u_7$-$u_2$)}
	\end{subfigure}
	\caption{Decomposition of FHR-UC data (a) via MEMD (b) and MVMD (c). The time axis of the plots for different modes (or IMFs) have been chosen to highlight the mode-mixing and mode-alignment property of MVMD as compared to BEMD.}
	\label{fig:fhr}
\end{figure}

Figure \ref{fig:fhr} shows the plots of selected decomposed modes obtained by applying the BEMD and MVMD to one of the patients' CTG record i.e., FHR-UC bivariate data; the total number of modes (or IMFs) obtained in both cases were $M=7$. Looking at the BEMD decomposition of the data, multiple instances of mode-mixing can be observed, for instance, at the time=20 sec in IMF1, IMF3 and IMF4; and at the time=100 sec in IMF3, IMF5 and IMF6 and at the time instant of around 65 seconds in IMF2 and IMF3. MVMD modes, on the other hand, are free of any artifacts arising due to mode-mixing in that each mode consists of narrow band frequencies. MVMD modes can also be seen to exhibit mode-alignment across all channels of the selected modes in the Figure \ref{fig:fhr}. Given that mode-mixing and misalignment of frequency information across channels can severely limit the performance of data driven multiscale algorithms, MVMD is expected to perform better than multivariate extensions of EMD across a range of practical applications.

\section{Conclusion}
We have proposed a novel extension of the variational mode decomposition (VMD) algorithm to multivariate data. VMD, in its original construction, aims to decompose a multicomponent single-channel signal into univariate modulated oscillations exhibiting limited bandwidth across a center frequency. That is achieved by formulating and solving a convex optimization problem that minimizes sum of bandwidths of all modes; bandwidth of a mode is estimated in VMD through squared $H_1$ norm of corresponding analytic signal, shifted to baseband by harmonic mixing with a complex exponential of center frequency estimate. We have proposed a generic extension of the VMD's variational model to be applied for multivariate data. The crux of the proposed method is to decompose multivariate data into its inherent \textit{multivariate modulated oscillations}. We have first defined a model for multivariate modulated oscillations that is based on constructing a Hilbert transformed multivariate analytic signal that has a \textit{common or joint frequency component} across all its channels. Based on that model we formulate a convex optimization problem that aims to minimize the sum of bandwidths of all modes across all channels while fulfilling multiple constraints of exact signal reconstruction across all data channels. The resulting estimates of multivariate signal modes along with their center frequencies have been shown to fulfill the narrow band requirements typically associated with EMD-inspired methods like VMD and synchrosqueezed transform. 

All multiscale data driven approaches for multivariate data require alignment of similar frequency content in a single scale or mode across all data channels, the so called mode-alignment property. The proposed method has been shown to exhibit this property on a variety of input multivariate signals ranging from test synthetic signal consisting of mixture of tones to multivariate wGn data and finally on data obtained from real world applications such as electroencephalogram (EEG) and cardiotocographic traces. As a result, we have demonstrated initial potential of the proposed method in data fusion application.

Being a generic extension of the VMD method, our proposed approach inherits all the advantages and flaws of the original VMD. Some of the challenges in the original VMD approach include: i) need to give predefined number of modes $K$ a priori; ii) inability to separate dc component in original data; and iii) problem to deal with signals exhibiting high level of nonstationarity such as sudden signal onset. We observed similar limitations in our proposed method. However, given the tremendous interest shown in VMD over the last few years since its inception, novel solutions to these shortcomings are expected which will improve both VMD and our proposed multivariate extension.



\begin{thebibliography}{1}

\bibitem{bib:stft} D. Gabor, ``Theory of communications'', \textit{Journal on IEE}, vol. 93, pp. 429-457, 1946.

\bibitem{bib:wavelet} I. Daubechies, ``Orthonormal basis of commonly supported wavelets'', \textit{Communications on Pure and Applied Mathematics}, vol. 41, pp. 909-996, 1988. 

\bibitem{bib:emd}N. E. Huang, Z. Shen, S. R. Long, M. L. Wu, H. H. Shih, Z. Quanan, N. C. Yen, C. C. Tung, and H. H. Liu, ``The empirical mode decomposition and the Hilbert spectrum for nonlinear and non-stationary time series analysis", \emph{Proceedings of the Royal Society A}, vol. 454, pp. 903--995, 1998.

\bibitem{bib:bio} C.~Park, D.~Looney, N.~Rehman and D.~P.~Mandic, ``Motor Imagery Signal Classification using Multivariate Empirical Mode Decomposition'', \emph{IEEE Transactions on Neural Systems and Rehabilitation Engineering}, vol.~21, no.~1, pp.~10-22,~2013.

\bibitem{bib:vag} S. Saleem, S. S. Naqvi, T. Manzoor, N. Rehman and J. Mirza, ``A Strategy for Classification of Vaginal vs. Cesarean Section Delivery: Bivariate Empirical Mode Decomposition of Cardiotocographic Recordings'', \emph{Frontiers in Physiology}, vol. 10, pp. 46, 2019. 

\bibitem{bib:clim} C. Franzke, ``Multi-scale analysis of teleconnection indices: climate noise and nonlinear trend analysis'', \emph{Nonlinear Processes in Geo Physics}, vol. 16, 65-76, 2009.

\bibitem{bib:mach} Y. Lei, J. Lin, Z. He, and M. J. Zuo, ``A review on empirical mode decomposition in fault diagnosis of rotating machinery'', \emph{Mechanical Systems and Signal Processing}, vol. 35, pp. 108-126, 2013.

\bibitem{bib:emd_form1} S. Meignen and V. Perrier, ``A new formulation for empirical mode decomposition based on constraint optimization'', \textit{IEEE Signal Processing Letters}, vol. 14, pp. 932-935, 2007.

\bibitem{bib:emd_form2} N. Pustelnik, P. Borgnat and P. Flandrin, ``Empirical mode decomposition revisited by multicomponent non-smooth convex optimization,'' \textit{Signal Processing}, vol. 102, pp. 313-331, 2014.

\bibitem{bib:sst} I. Daubechies, J. Lu, and H. T. Wu, ``Synchrosqueezed wavelet transforms: An Empirical Mode Decomposition-like tool,'' \emph{Applied Computational Harmonic Analysis}, vol. 30, no. 2, pp. 243–261, 2011.

\bibitem{bib:ewt} J. Giles, ``Empirical wavelet transform,'' \emph{IEEE Transactions on Signal Processing}, vol. 61, pp. 3999-4010, 2013.

\bibitem{bib:reassign} F. Auger and P. Flandrin, ``Improving the readability of time-frequency and time-scale representations by the reassignment method'', \emph{IEEE Transactions on Signal Processing}, vol. 43, no. 5, pp. 1068–1089, 1995.

\bibitem{bib:vmd} K. Dragomiretskiy and D. Zosso, ``Variational Mode Decomposition'', \emph{IEEE Transactions on Signal Processing}, vol. 62, no. 3, pp. 531-544, 2014.

\bibitem{bib:memd_seizure} A. Zahra, N. Kanwal, N. Rehman, S. Ehsan and K. D. McDonald-Maier, ``Seizure detection from EEG signals using multivariate empirical mode decomposition'', \emph{Computers in Biology and Medicine}, vol. 88, pp. 132-141, 2017. 

\bibitem{bib:memd_ecg} G. Han, B. Lin and Z. Xu, ``Electrocardiogram signal denoising based on empirical mode decomposition technique: an overview'', \emph{Journal of Instrumentation}, vol. 12, no. 3. pp. P03010-P03010, 2017.

\bibitem{bib:memd_den} H. Hao, H.L. Wang and N. Rehman, ``A joint framework for multivariate signal denoising using multivariate empirical mode decomposition'', \emph{Signal Processing}, vol. 135, pp. 263-273, 2017. 

\bibitem{bib:memd} N.~Rehman and D.~P.~Mandic, ``Multivariate Empirical Mode Decomposition'', \emph{Proceedings of the Royal Society A}, vol.~466, no.~2117, pp.~1291-1302,~2010.

\bibitem{bib:bemd} G. Rilling, P. Flandrin, P. Goncalves, J. M. Lilly ``Bivariate Empirical Mode Decomposition'', \emph{IEEE Signal Processing Letters}, vol.~14, pp.~936-939,~2007.

\bibitem{bib:temd} N.~Rehman and D.~P.~Mandic, ``Empirical Mode Decomposition for Trivariate Signals'', \emph{IEEE Transactions on Signal Processing}, vol.~58, no.~3, pp.~1059-1068,~2010. 

\bibitem{bib:msst} A. Ahrabian, D. Looney, L. Stankovic, D. P. Mandic, ``Synchrosqueezing-based time-frequency analysis of multivariate data'', \emph{Signal Processing}, vol. 106, pp. 331-341, 2015.

\bibitem{bib:mewt} O. Singh, R. K. Sunkaria, ``An empirical wavelet transform based approach for multivariate data processing application to cardiovascular physiological signals'', \emph{Bio-Algorithms and Med-Systems}, vol. 14, no. 4, 2018.

\bibitem{bib:cvmd} Y. Wang, F. Liu, Z. Jiang, S. He and Q. Mo, ``Complex variational mode decomposition for signal processing applications'', \emph{Mechanical systems and signal processing}, vol. 86, pp. 75-85, 2018.

\bibitem{bib:fb_memd} N.~Rehman and D.~P.~Mandic, ``Filter Bank Property of Multivariate Empirical Mode Decomposition'', \emph{IEEE Transactions on Signal Processing}, vol.~59, no.~5, pp.~2421-2426,~2011.

\bibitem{bib:admm} S. Boyd, N. Parikh, E. Chu, B. Peleato and J. Eckstein, ``Distributed Optimization and Statistical Learning via the Alternating Direction Method of Multipliers'', \emph{Foundations and Trends in Machine Learning}, vol. 3, pp. 1-122, 2011.

\bibitem{bib:Abdullah} U. Abdullah, N. Rehman, M. Khan, and D. P. Mandic, ``A multivariate EMD based approach to Pan-sharpening'', \textit{IEEE Transactions in Geoscience and Remote Sensing}, vol. 53, no. 7, pp. 3974-3984, 2015.

\bibitem{bib:lilly1} J. M. Lilly, ``Modulated Oscillations in Three Dimensions'', \emph{IEEE Transactions on Signal Processing}, vol. 59, no. 12, pp. 5930-5943, 2011.

\bibitem{bib:lilly2} J. M. Lilly, and S. C. Olhede, ``Analysis of Modulated Multivariate Oscillations,'' \emph{IEEE Transactions on Signal Processing}, vol. 60, pp. 600-612, 2012.

\bibitem{bib:lilly3} S. C. Olhede, ``Analysis of Modulated Multivariate Oscillations,'' \emph{Philosophical Transactions of the Royal Society A}, vol. 371, 2013.

\bibitem{bib:memd_fus} N. Rehman and D. Looney and T.M. Rutkowski and D. P. Mandic, "Bivariate {EMD}-based Image Fusion,"\emph{Proceedings of the 2009 IEEE Workshop on Statistical Signal Processing,
	Wales,} 2009.

\bibitem{bib:looney} D. Looney and D. P. Mandic, “Multiscale Image Fusion Using Complex Extensions of EMD”, \emph{IEEE Transactions On Signal Processing}, vol. 57, no. 4, April 2009.


\bibitem{bib:emd_fb} P. Flandrin, G. Rilling, P. Goncalves, ``Empirical mode decomposition as a filter bank'', \emph{IEEE Signal Processing Letters}, vol.~11, no.~2, pp.~12-14,~2004.

\bibitem{bib:memd_fb} N.~Rehman and D.~P.~Mandic, ``Filter Bank Property of Multivariate Empirical Mode Decomposition'', \emph{IEEE Transactions on Signal Processing}, vol.~59, no.~5, pp.~2421-2426,~2011.

\bibitem{bib:warrick} P. A. Warrick, E. F. Hamilton, D. Precup, and R. E. Kearney, ``Classification of normal and hypoxic fetuses from systems modeling of intrapartum cardiotocography'' \emph{IEEE Transactions on Biomedical Engineering}, vol. 57, pp. 771–779, 2010.

\bibitem{bib:saleem} S. Saleem, S. S. Naqvi, T. Manzoor, N. Rehman and J. Mirza, ``A Strategy for Classification of Vaginal vs. Cesarean Section Delivery: Bivariate Empirical Mode Decomposition of Cardiotocographic Recordings'', \emph{Frontiers in Physiology}, vol. 10, pp. 46, 2019. 

\bibitem{bib:goldberger} A. L. Goldberger \textit{et. al.} ``PhysioBank, PhysioToolkit, and PhysioNet: components of a new research resource for complex physiologic signals'', \emph{Circulation 101}, e215–e220. doi: 10.1161/01.CIR.101.23.e215, 2000.

\end{thebibliography}
\end{document}